\documentclass[aps,prc,twocolumn,showpacs,superscriptaddress,floatfix]{revtex4}
\usepackage{graphicx}
\usepackage{longtable}

\begin{document}

\title{Shape coexistence in $^{153}$Ho}

 \author {Dibyadyuti  Pramanik}

 \affiliation{Indian   Institute   of   Engineering  Science  and
Technology, Shibpur, Howrah - 711103, INDIA}

 \author {S. Sarkar}
\thanks{ss@physics.iiests.ac.in}
 \affiliation{Indian   Institute   of   Engineering  Science  and
Technology, Shibpur, Howrah - 711103, INDIA}

  \author {M.  Saha Sarkar}
  \thanks{ maitrayee.sahasarkar@saha.ac.in}

\affiliation{Saha  Institute  of  Nuclear  Physics,  Bidhannagar,
Kolkata - 700064, INDIA}

 \author{Abhijit Bisoi}

  \affiliation{Indian   Institute   of  Engineering  Science  and
Technology, Shibpur, Howrah - 711103, INDIA}

 \author{Sudatta  Ray}\thanks{Amity  University,  Noida - 201303,
INDIA}   \affiliation{Saha   Institute   of   Nuclear    Physics,
Bidhannagar, Kolkata - 700064, INDIA}

\author{Shinjinee    Dasgupta}\thanks{Heritage    Institute    of
Technology, Anandapur, Kolkata – 700107, INDIA} \affiliation{Saha
Institute of Nuclear Physics, Bidhannagar,
Kolkata - 700064, INDIA}

 \author{ A.Chakraborty} \affiliation{Visva-Bharati, Santiniketan
- 731235, INDIA}

\author{Krishichayan} \affiliation{Duke University, Durham, North
Carolina, USA}

 \author{Ritesh      Kshetri}     \affiliation{Sidho-Kanho-Birsha
University, Purulia - 723101, INDIA}

 \author{Indrani  Ray}  \affiliation{Saha  Institute  of  Nuclear
Physics, Bidhannagar, Kolkata - 700064, INDIA}

 \author{S.  Ganguly}  \affiliation{Bethune  College,  Kolkata  -
700006, INDIA}

 \author{M.  K.  Pradhan} \affiliation {Saha Institute of Nuclear
Physics, Bidhannagar, Kolkata - 700064, INDIA}

\author{M. Ray Basu} \affiliation{University of Calcutta, Kolkata
- 700073, INDIA}

\author{R.  Raut} \affiliation {UGC-DAE Consortium for Scientific
Research, Kolkata- 700098, INDIA}

\author{G. Ganguly} \affiliation {University of Calcutta, Kolkata
- 700073, INDIA}

\author{S.S.   Ghugre}   \affiliation   {UGC-DAE  Consortium  for
Scientific Research, Kolkata- 700098, INDIA}

\author{A.K.   Sinha}   \affiliation   {UGC-DAE   Consortium  for
Scientific Research, Kolkata- 700098, INDIA}

\author{S.K.   Basu}   \affiliation  {Variable  Energy  Cyclotron
Centre, Kolkata - 700064, INDIA}

 \author{S. Bhattacharya} \affiliation {Saha Institute of Nuclear
Physics, Bidhannagar, Kolkata - 700064, INDIA}

  \author{A.  Mukherjee}  \affiliation{Saha  Institute of Nuclear
Physics, Bidhannagar, Kolkata - 700064, INDIA}

 \author{P.  Banerjee}  \affiliation{Saha  Institute  of  Nuclear
Physics, Bidhannagar, Kolkata - 700064, INDIA}

\author{A.  Goswami}  \affiliation  {Saha  Institute  of  Nuclear
Physics, Bidhannagar, Kolkata - 700064, INDIA}

\date{\today}

\begin{abstract}
The   high-spin  states  in  $^{153}$Ho,  have  been  studied  by
$^{139}_{57}$La($^{20}$Ne, 6n) reaction at a projectile energy of
139 MeV at Variable  Energy  Cyclotron  Centre  (VECC),  Kolkata,
India,  utilizing  an  earlier  campaign of Indian National Gamma
Array  (INGA)   setup.   Data   from   gamma-gamma   coincidence,
directional  correlation  and polarization measurements have been
analyzed to assign and confirm the  spins  and  parities  of  the
levels.  We  have  suggested a few additions and revisions of the
reported level scheme of $^{153}$Ho. The RF-gamma time difference
spectra have been useful to confirm the half-life of an isomer in
this nucleus. From the comparison of experimental and theoretical
results, it is found that there are definite indications of shape
coexistence in this  nucleus.  The  experimental  and  calculated
lifetimes  of  several  isomers  have been compared to follow the
coexistence and evolution of shape with increasing spin.

\end{abstract}


\maketitle

\section{Introduction}

       The  neutron  deficient rare-earth isotopes near the magic
nucleus $^{146}$Gd have shown multitude of structural features as
functions of neutron number as well  as  spin  \cite{nndc,shape}.
For isotones with N = 86, excitation spectra show single particle
nature  associated  with non-collective modes. For N = 88, strong
collectivity in  terms  of  appreciable  prolate  deformation  is
manifested in the low-lying spectra. However, even for the nuclei
which are very close to $^{146}$Gd, although low spin excitations
are usually very irregular and complex indicating spherical shape
with  single  or multi-particle excitations, at relatively higher
energies superdeformed (SD) bands  \cite{nndc,dy152a,dy152b}  are
observed.  The  observations  indicate that these nuclei are very
soft against shape changes. The features  have  been  interpreted
theoretically   using   microscopic  shell-model  and  mean-field
descriptions \cite{shape}. One of the distinguishing features  of
this  mass  region  is  the  existence  of an island of high spin
isomers \cite{nndc,iso,iso1}, which  are  excited  in  heavy  ion
reactions.  These  isomers  can  also  indicate a sharp change of
structural    configurations    within    the    same     nucleus
\cite{iso,iso1}.

   This   mass   region   has   been   investigated   extensively
particularly for even - even  nuclei  \cite{nndc}.  It  has  been
possible  to  observe  structural  changes  and shape coexistence
effects  for   them.   Some   nuclei   in   this   region,   like
$^{152}_{66}$Dy$_{86}$   \cite{dy152a,dy152b},   are  found  to  be
near-spherical oblate  in  shape  at  low  spins.  However,  they
exhibit     coexisting    collective    prolate    shapes    like
superdeformation at higher spins. On the  other  hand  there  are
evidences  of  excitation pattern in $^{154}_{66}$Dy$_{88}$ similar
to collective prolate rotors at low spins  which  evolve  to  non
collective oblate shape at high spins. Nuclei with higher neutron
numbers   (N   $>90$),  behave  like  collective  prolate  rotors
throughout their entire excitation pattern.  For  odd  Z  nuclei,
like  Holmium  (Ho),  although  individual  studies  of different
isotopes      \cite{ho151,ho152a,ho152b,153hon,ho154}      exist,
systematic  analysis  of the structural evolution of this element
with variations in neutron number is scarce.  Ho  (Z=67)  is  the
nearest  odd-Z neighbor of most extensively studied Dy (Z=66). A
systematic  study  \cite{dey,anagha,db1,db2,db3,db4,db5}  of  the
isotopes  of  Ho  element  has  been  initiated to understand how
nuclear structure  differs  due  to  the  addition  of  a  single
unpaired proton. In the present work, $^{153}$Ho isotope has been
studied.

The  neutron  deficient isotope of Ho, $^{153}$Ho has an unpaired
proton coupled to the even-even core $^{152}$Dy. $^{153}$Ho with
neutron    number    N=86    has    been    studied    previously
\cite{153hon,153ho,153hotri} to high spins  (J  $\simeq$  $81/2$)
\cite{153ho}.   Its   $11/2^-$  ground  state  quadrupole  moment
measured  using  Laser  resonance  ionization  mass  spectroscopy
(LRIMS) method is negative indicating an oblate shape \cite{mom}.
Even   in   the   latest  compilation  \cite{153hon},  there  are
uncertainties in the spin and parity assignments of  the  excited
levels.  In  an  earlier work \cite{153ho}, the structure of this
nucleus has been analyzed phenomenologically in  terms  of  shell
model  configurations  of  the  excited  states.  There have been
suggestions that the $3^-$ octupole vibration may  play  a  major
role   in   its   structure   \cite{153ho}.   In  a  later  study
\cite{153hotri}, a low deformation collective band in  $^{153}$Ho
has  been found to be associated with a triaxial shape with large
positive value triaxiality parameter ($\gamma$).

In   the  present  work,  the  high-spin  states  in  $^{153}$Ho,
populated in a heavy ion reaction have been  studied.  Data  from
gamma-gamma coincidence, directional correlation and polarization
measurements  have  been analyzed to assign and confirm the spins
and parities of the levels. A few additions and revisions of  the
reported  \cite{153ho}  level  scheme  of  $^{153}$Ho  have  been
suggested. The RF-gamma time difference spectra have been used to
confirm the half-life of a nanosecond  isomer  in  this  nucleus.
Total  Routhian  Surface  (TRS)  and  Particle  Rotor Model (PRM)
calculations have also been performed.  From  the  comparison  of
experimental  and  theoretical  results in the entire spin-energy
domain, it is demonstrated that there are definite indications of
shape coexistence in this nucleus. The comparison of experimental
and calculated lifetimes  of  several  isomers  observed  in  the
excitation  spectra  of this nucleus helps in following the shape
coexistence and its evolution with increasing spins.

\begin{figure}
\vspace{2.5cm}
\includegraphics[width=\columnwidth,angle=0]{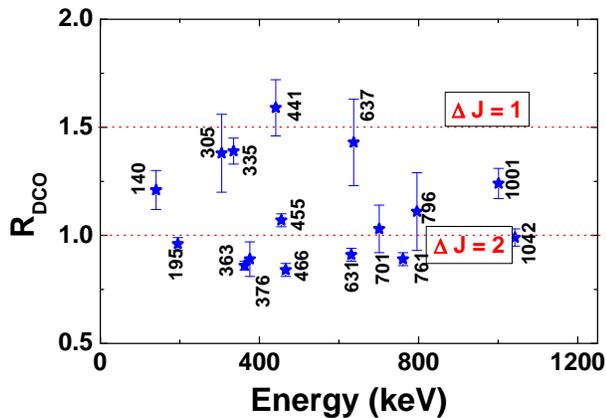}
\vspace{-3cm}
\caption{\label{dco}(Color   Online)   Experimental   DCO  ratios
($R_{DCO}$) for few transitions in $^{153}$Ho. Typical values of
$R_{DCO}$s determined for $\Delta J=0$ and 1 transitions from E2 
gated  spectra are indicated in the figure by dotted lines.}

    \end{figure}

\section{Experimental Details and Data  Analysis}

High  spin  states of $^{153}$Ho were populated by bombarding 139
MeV $^{20}Ne$ beam on  a  La$_2$O$_3$  target  at  Variable  Energy
Cyclotron  Centre  (VECC),  Kolkata.  The  target of thickness $3
mg/cm^2$ was prepared by centrifuge method on Al backing of  $2.2
mg/cm^2$.  $^{153}$Ho  was  populated  most  strongly, along with
population of $^{151,152,154}$Ho, $^{152-153}$Dy nuclei.  Results
from  preliminary  data  analysis  from this experiment have been
reported earlier \cite{dey,db1,db2,db3}. The experiment has  been
carried  out  using  one  of the earlier campaigns \cite{raut} of
Indian National Gamma Array (INGA) setup, which comprised of  six
Compton-suppressed Clover detectors. In this setup, the detectors
were  placed  at $40^o$(2), $90^o$(2), $125^o$(2) with respect to
the beam direction. Data were acquired in LIST mode. At least two
correlated  gamma  energies  from   the   Clovers,   their   time
information  as  well  as  corresponding RF time information with
respect to the master gate have been included in the LIST data.

The  energy  and  efficiency  calibration of the Clover detectors
have  been  done  using  radioactive  $^{133}$Ba  and  $^{152}$Eu
sources.  The  coincidence  events were sorted into two symmetric
$\gamma$-$\gamma$ matrices with time gates of 800 ns  and  200ns.
These matrices are used to generate various background subtracted
gated  spectra.  The  matrices  were  analyzed  using  the  codes
INGASORT \cite{ingasort} and RADWARE \cite{radware}. To  generate
background  subtracted  gated spectra, INGASORT program has been
used. In this program, the background is eliminated, on  peak  by
peak basis.

\subsection{Angular Correlation Data Analysis}

Angle  dependent  asymmetric $\gamma$-$\gamma$ matrices have been
generated  to  determine  the  multipolarity  of  $\gamma$-   ray
transitions   from  directional  correlation  of  $\gamma$-  rays
emitted from excited oriented states (DCO) measurements. The  DCO
ratio  ($R_{DCO}$)  of  a  $\gamma$  transition ($\gamma_{1})$ is
defined  as  the  ratio  of  intensities  of  that   $\gamma$-ray
($I_{\gamma_{1}}$)  for  two different angles in coincidence with
another $\gamma$-ray ($\gamma_{2})$ of known multipolarity. It is
given by the ratio, defined as

\begin{eqnarray}
R_{DCO} = \frac{I_{{\gamma}_1}  ~\text{observed at}~ 40^{\circ},~
\text{gated by}~ \gamma_2~ \text{at}~ 90^{\circ}~} { I_ {{\gamma}_1}
~\text{observed at}~ 90^{\circ}, ~\text{gated by}~ \gamma _2~ \text{at}~
40^{\circ} }\nonumber
    \end{eqnarray}

The  DCO  ratio  of  each $\gamma$ has been obtained by putting a
gate on a $\gamma$ transition of known multipolarity with zero or
very small mixing  ratio  (Fig.  \ref{dco}).  For  the  stretched
transitions  with  same  multipolarity  as the gating transition,
$R_{DCO}$ value should be very  close  to  unity.  For  different
multipolarities  of  the  gating  and  projected transitions, the
$R_{DCO}$ value depends on the angle between  the  detectors  and
the   amount   of  mixing  present  in  the  mixed  multipolarity
transition. For the assignment of spins  and  the  $\gamma$  -ray
multipole  mixing  ratios ($\delta$), the experimental DCO values
were compared with the theoretical values calculated by using the
computer code  ANGCOR  \cite{macias}.  Spin  alignment  parameter
$\sigma$/J  =  0.3 was used for this calculation. This choice was
guided  by  several  earlier   works   \cite{30p,34cl,33s,142sm}.
However,   the   choice   has   been  also  tested  further.  The
experimental mixing ratio of 140  keV  gamma  determined  in  the
present  work  has  been  used  to calculate the K-shell internal
conversion coefficient from BrIcc  v2.3S  Conversion  Coefficient
Calculator  \cite{bricc}  which  agrees  reasonably well with the
experimental K-shell electron  conversion  coefficient  for  this
transition as reported in Fig. 2 of the reference \cite{153ho} as
well as in \cite{153hon}.

\begin{figure}\vspace{2.5cm}

\includegraphics[width=\columnwidth,angle=0]{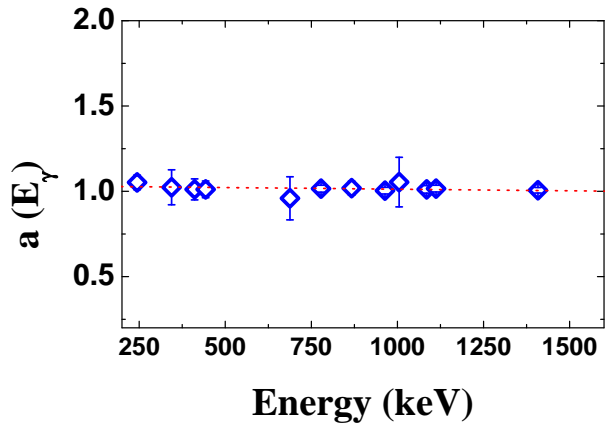}
\vspace{-2.5cm}
\caption{\label{asym}(Color     Online)    Asymmetry    parameter
$a(E_\gamma)$ plotted as a function of energy. The fitted straight
line is shown as dotted line in the figure.}
    \end{figure}

\subsection{Polarization Data Analysis}

Integrated  polarization asymmetry measurements (IPDCO) have been
done to determine the electric or magnetic nature of the $\gamma$
-ray transitions. Two asymmetric IPDCO matrices were  constructed
from  the  data.  The  first  (second)  matrix  named as parallel
(perpendicular)  was  constructed  having  on  first   axis   the
simultaneous  events  recorded  in the two crystals of the 90$^o$
Clover  detector  which  are  parallel  (perpendicular)  to   the
emission plane and on the second axis the coincident $\gamma$ ray
registered  in  any other detector. The polarization asymmetry is
defined as

\begin{equation}
\Delta_{IPDCO}=\frac{a(E_\gamma)N_\perp-N_\parallel}
{a(E_\gamma)N_\perp+N_\parallel}
 \end{equation}

where $N_\perp$ and $N_\parallel$ are the intensities of the full
energy peaks observed in the perpendicular and parallel matrices,
respectively. The correction term a($E_\gamma$) is introduced due
to  asymmetry  in  the  response of the different crystals of the
Clover detector at ${90}^o$. It is defined as

\begin{equation}
a(E_\gamma)=\frac{N_\parallel(unpolarized)}{N_\perp(unpolarized)}.
\end{equation}

In the present experiment $a$ is measured as a function of energy
of  unpolarized $\gamma$-rays from radioactive $^{152}$Eu source.
Fig.\ref{asym} shows the variation of $a$ with $E_\gamma$ and  it
was  fitted  with  the  expression $a$($E_\gamma$) = $a_0$ + $a_1
E_\gamma$ resulting into  $a_0$  =  1.03  and  $a_1  \sim$  -1.93
$\times 10^{-5}$, where $E_\gamma$ is in keV.

\begin{figure}\vspace{2.5cm}

\includegraphics[width=\columnwidth,angle=0]{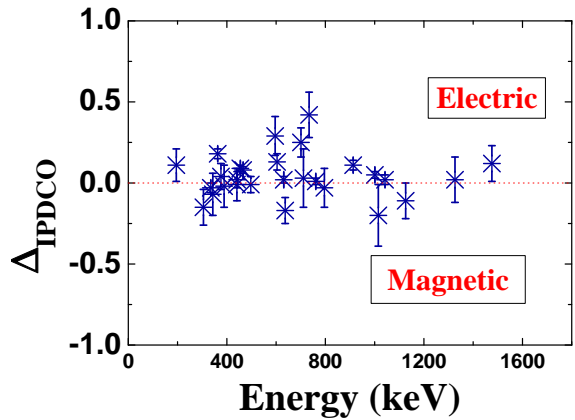}
\vspace{-2.5cm}
\caption{\label{pdco}(Color  Online)  Polarization symmetry for a
few transitions in $^{153}$Ho. The dotted line separates the regions
corresponding to pure Electric and Magnetic transitions. }
    \end{figure}

For  determination  of  experimental  polarization asymmetry from
each of the IPDCO matrices, we have put gates on $\gamma$s on the
second axis and observed the projected parallel and perpendicular
spectra of the 90$^o$  Clover  detectors  (Fig.  \ref{pdco}).  A
positive  (negative)  value of $\Delta_{IPDCO}$, indicates a pure
electric  (magnetic)  transition.  But  for  mixed   transitions,
usually this value is close to zero and the sign varies depending
on the extent of mixing.

\begin{figure*}\vspace{2.5cm}

\centering
  \begin{tabular}{c@{}@{}c}
\includegraphics[width=.4\textwidth, height=0.38\textwidth,angle=0]{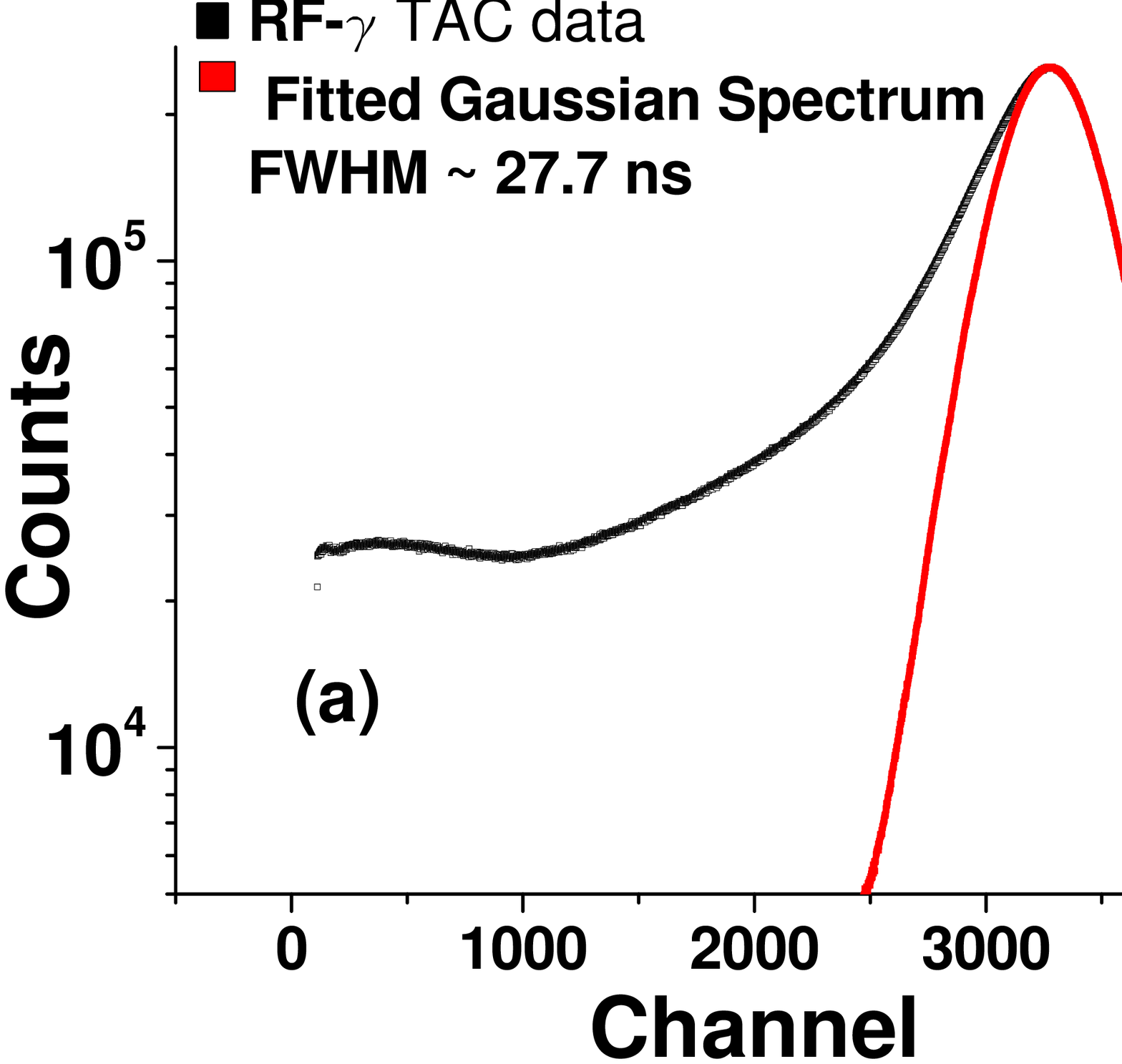} &
    \includegraphics[width=.4\textwidth, height=0.38\textwidth,angle=0]{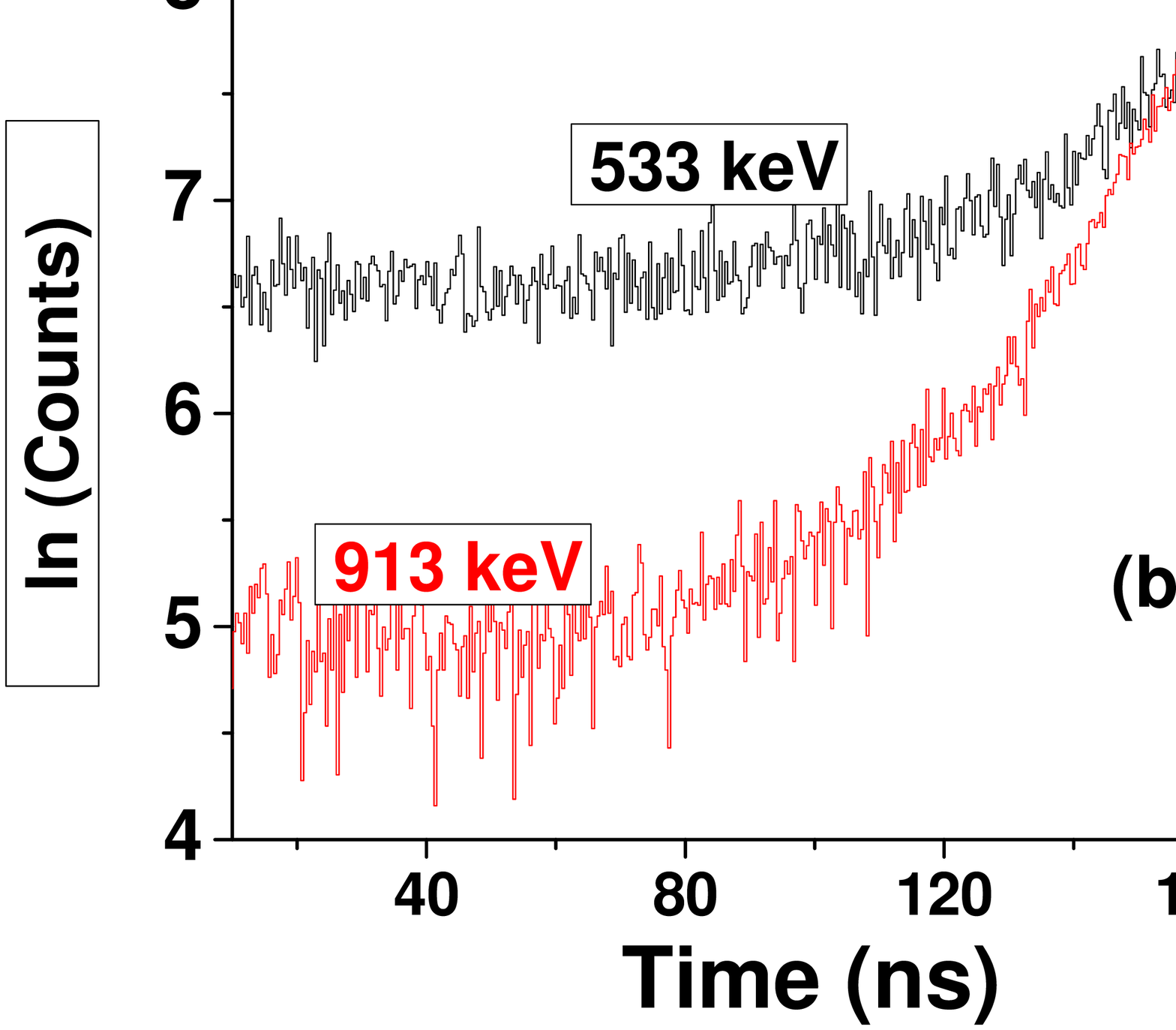}\\
\end{tabular}
 \vspace{-2.5cm}
\caption{
\label{rftac}(Color  Online) (a) The RF-$\gamma$ TAC spectrum without
any  gate  on  $\gamma$  energies(left).  (b) (Right)  Comparison  of
RF-$\gamma$  TAC  spectra  gated by 913 keV prompt transition and
533 keV delayed transition decaying from the isomer at  2772  keV
through 36 keV. The logarithmic (ln) values of counts  (to the base e)
are plotted against time (ns) for ease of comparison.  } \end{figure*}

\subsection{Lifetime data analysis}

In  this experiment, the RF frequency is $\simeq$5.56 MHz, so the
time difference  between  two  consecutive  RF  pulses  from  the
cyclotron   is   around   180  ns.  Fig.  \ref{rftac}  shows  the
RF-$\gamma$ time-difference to amplitude converted (TAC) spectrum
with $\gamma$ energies ranging from $\approx$  100  keV  to  4000
keV.  This  TAC spectrum has been taken within a range of 200 ns.
In this condition the resolution  of  the  prompt  time  spectrum
comes  out  to  be  $\simeq$  27.7  ns without any restriction in
energy.

\section{\bf{RESULTS}}

\subsection{Level Scheme}

\begin{figure*}
\includegraphics[width=1.5\columnwidth,angle=0]{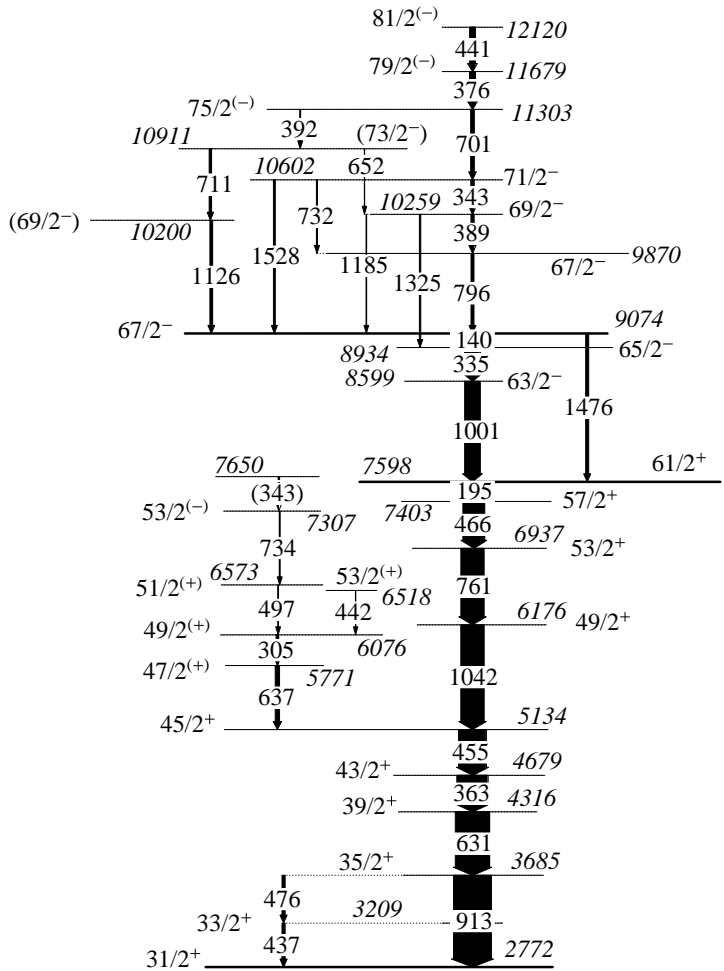}
\vspace{-4 cm}
\caption{\label{153ho-level}  Partial  level  scheme of $^{153}$Ho}.
\end{figure*}

\begin{figure*}
\hspace{-2cm}
\vspace{1cm}
\includegraphics[width=1.5\columnwidth,height=2\columnwidth,angle=-90]{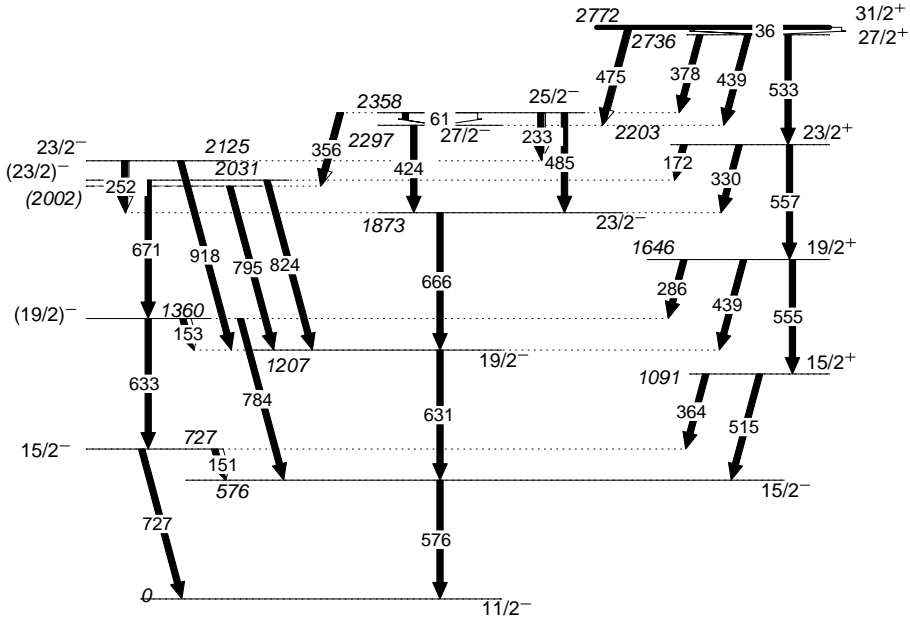}
\vspace{-2cm}
\caption{\label{153ho-below}  Partial  level scheme of $^{153}$Ho
below the isomer $31/2^+$ at 2772 keV}. \end{figure*}

The  level scheme of $^{153}$Ho, shown in Fig. \ref{153ho-level},
has been established using the coincidence relationship, relative
intensities, $R_{DCO}$ and $\Delta_{IPDCO}$  ratios  of  $\gamma$
rays.  All  transitions  above  2772  keV  ($31/2^+$) up to $E_x$
$\sim$ 12 MeV and a tentative spin of $81/2^-$  reported  in  the
earlier  work  by  Radford  {\it  et al.} \cite{153ho}, have been
observed in the present experiment.  The  transitions  below  the
long  -  lived  isomer  at  2772  keV have also been observed and
confirmed in this work (Fig. \ref{153ho-below}). However, in most
of  the  cases  their  intensities  could   not   be   determined
unambiguously.  Thus the figure (Fig. \ref{153ho-below}) does not
contain the intensities of these transitions.

Symmetric  $\gamma$  -  $\gamma$  matrix  has  been used to place
different gamma transitions in the level scheme.  To  assign  the
spins  and  parities  of  the  levels,  the  conventional DCO and
polarization  measurements  have  been  performed.  The  relative
intensities  of these transitions have been obtained from 913-keV
gated spectrum  and  relative  intensities  are  normalized  with
respect  to  the  intensity  of  363 keV transition. The relative
intensities, experimental $R_{DCO}$ values,  mixing  ratios  (for
mixed transitions) and experimental polarization asymmetry values
are  listed  in  Table \ref{intense}. For transitions parallel to
913  keV  transition,  alternate  gating  transition   has   been
considered to get intensities.

The  intensities  of  transitions  below the long-lived isomer at
2772 keV, are also determined from  200  ns  time  gated  matrix.
However,  the ratios provided in Table \ref{rftact} indicate that
their intensities should increase by $\simeq 30\%$ if  they  were
measured  from  a  matrix  generated  with  800  ns time gate. As
expected, the angular correlation data for transitions below  the
isomer  have been found to be isotropic with gates on transitions
above the isomer. The intensities of gamma rays below the  isomer
in  asymmetric  matrix  were  quite low. Moreover, the excitation
pattern is fragmented and contains many  gamma  rays  of  similar
energies  as  those above the isomer. So it was also not possible
to do correlation measurements of these transitions with gates on
transitions below the isomer.

The  relative intensities of most of the gamma rays in $^{153}$Ho
up to that of 466 keV gamma ray  emitted  from  7403  keV  level,
determined  from  the  spectra gated by 913 keV gamma ray emitted
from 2772 keV level, show good agreement with placements shown in
Ref. \cite {153ho}. However, the positions of 335  and  1001  keV
gammas   have   been  interchanged  in  the  level  scheme  (Fig.
\ref{153ho-level})   based   on    their    intensities    (Table
\ref{intense}).  The  suggestion  for reversal of the ordering of
the 761- and 1042-keV transitions in Ref. \cite{153hotri} is  not
supported by the intensities of these transitions obtained in the
present  work.  For  low  energy  gammas  like,  195 - , 140- keV
transitions inclusion of  internal  conversion  (IC)  corrections
have  been  important  for  proper placement. We have utilized IC
coefficients  obtained  from  online  ICC  calculator  of   BRICC
\cite{bricc}.

Several  isomers  have been reported in $^{153}$Ho in the earlier
works  \cite{153hon,153ho}.  The  relative  intensities  in   the
present  work  also  provide  information  regarding the existing
isomers. As two gamma transitions of 631 and 633 keV exist in the
level scheme above and below the 2772  keV  isomeric  level,  the
intensities  in Table \ref{intense} have been quoted with respect
to that of 363 keV transition. Therefore any fall in intensity of
gammas decaying from levels above the previously reported $\simeq
500 ps$ isomer at 4679 keV level could not be observed. Moreover,
this half life is also not  comparable  to  200ns  prompt  window
selected  for the time-gated symmetric matrix to be manifested in
the  decrease  of  intensity.  For  the  reported  3  ns   isomer
\cite{153ho} at 7598 keV, the relative intensity of 195 keV gamma
transition decaying from it is $\simeq 75 \%$ compared to $\simeq
52\%$  intensity  of  the  1001 keV gamma transition feeding this
level, supporting the possibility of an isomer. Similarly, it  is
found  that  (Table  \ref{intense})  the  feeding to the 9074 keV
state is highly fragmented. The several  weak  gamma  transitions
(796  keV,  1126 keV, 1185 keV and 1528 keV) feed this level. The
level sequence above this state is also highly irregular. In  the
earlier work a $\simeq 300 ps$ isomer was reported at this level.
This  sudden  change  in  excitation  pattern  can  also indicate
presence of a structure isomer at this level.

For  most  of  the  gamma  rays, the DCO and polarization results
satisfy earlier assignments  \cite{nndc,153ho}.  However,  for  a
few,  some differences have been noted. In Ref. \cite{153ho}, 455
keV transition has been indicated as a  M1  transition.  The  DCO
measurement  in  the  present  work (Fig. \ref{dco}) results to a
value of $R_{DCO}$ = 1.07(3) from $90^o~\rm{vs}~40^o$  asymmetric
matrix  when gated by an E2 transition. The theoretical $R_{DCO}$
values have been  calculated  by  varying  the  mixing  ratio  to
reproduce the experimental DCO ratio. This comparison indicated a
large mixing ratio (E2/M1) $(\delta)$ 0.33(3) for this transition
(Table \ref{intense}).

According to the present data analysis (Table \ref{intense}), 637
keV  gamma  transition  is  M1  in  nature  and  the mixing ratio
($\delta$)  is  0.08,  whereas  it   was   previously   mentioned
\cite{153ho,nndc}  as  an  E1  transition.  The  spin  and parity
assignments were not mentioned \cite{153ho,nndc}  for  701,  376,
441  keV transitions. In the present work, it has been found that
701 and 376 keV transitions are of E2  character  while  441  keV
transition is M1 in nature. Previously, nature of 734 and 711 keV
transitions  were not specified \cite{153ho}. Present measurement
indicates that 734 keV and 711 keV transitions  have  E1  and  E2
character, respectively (Table \ref{intense}).

\begin{table*}
\caption{\label{intense}    Relative    Intensity    ($I_{rel}$),
$R_{DCO}$, $\Delta_ {IPDCO}$ and the mixing ratio  ($\delta$)  of
the $\gamma$ transitions in $^{153}$Ho. }

\begin{tabular} {cccccccc}
\hline
\hline
$E_{\gamma}$(keV)& $J_i^{\pi}$& $J_f^{\pi}$&  $I_{rel}$\hfil&
$E_{gate}$(keV)& \hfil$R_{DCO}$\hfil& \hfil $\delta$ \hfil& \hfil
$\Delta_{IPDCO}$ \\

\hline
\hline
140& $67/2^-$& $65/2^-$& 32.72& 913[E2]& 1.21(9)& $0.20_{-0.05}^{+0.09}$ \\
195& $61/2^+$& $57/2^+$& 75.38& 913[E2]& 0.96(3)&  E2&0.11(1) \\
233\footnote{Below the isomer}& $25/2^-$& $23/2^-$& 6.80 \\
252$^a$& $23/2^-$& $23/2^-$& 4.66  \\
287$^a$& $19/2^+$& $(19/2)^-$& 4.82 \\
305& $49/2^{(+)}$& $47/2^{(+)}$& 7.97& 913[E2]& 1.38(8)& $0.11_{-0.05}^{+0.11}$& -0.15(1) \\
335& $65/2^-$& $63/2^-$& 48.45& 913[E2]& 1.39(6)&  $0.10(3)$&-0.03(3) \\
343& $71/2^-$& $69/2^-$& 11.05& 761[E2]& 1.24(3)& $0.19_{-0.08}^{+0.09}$&-0.07(3) \\
363& $43/2^+$& $39/2^+$& 100 & 913[E2]& 0.86(2)& E2&0.18(3)  \\
376& $79/2^{(-)}$& $75/2^{(-)}$& 19.12& 913[E2]& 0.89(8)& E2& 0.04(8) \\
378$^a$& $27/2^+$& $25/2^-$&22.17 \\
389& $69/2^-$& $67/2^-$& 9.68&  466[E2]& 1.27(4)& $0.17_{-0.08}^{+0.11}$&-0.02(3) \\
424$^a$& $27/2^-$& $23/2^-$& 5.91& \\
441\footnote{May contain contribution from 442 keV ($53/2^{(+)} \rightarrow 49/2^{(+)}$)}&
$81/2^{(-)}$& $79/2^{(-)}$& 18.04&  913[E2]& 1.59(3)&&-0.01(.10)\\
455& $45/2^+$& $43/2^+$& 88.68& 913[E2]& 1.07(3)& 0.33(3)& 0.09(2)\\
466& $57/2^+$& $53/2^+$& 71.06&  913[E2]&0.84(3)& E2&0.08(2)  \\
497& $51/2^{(+)}$& $49/2^{(+)}$& 2.44& 637[M1]& 0.79(4)& 0.3(5)&-0.01(5) \\
515$^a$& $15/2^+$& $15/2^-$& 13.87 \\
533$^a$& $27/2^+$& $23/2^+$& 33.12\\
557$^a$& $23/2^+$& $19/2^+$& 52.35 \\
576$^a$& $15/2^-$& $11/2^-$& 36.15 \\
631& $39/2^+$& $35/2^+$ &113.08&913[E2]& 0.91(3)& E2& 0.02(2) \\
637& $47/2^{(+)}$& $45/2^+$& 14.13& 913[E2]& 1.43(1)& $0.08_{-0.09}^{+0.12}$& -0.17(8)\\
666$^a$& $23/2^-$& $19/2^-$& 8.31 \\
701& $75/2^{(-)}$& $71/2^-$& 12.03& 913[E2]& 1.03(1)& E2&0.10(8) \\
711& $73/2^-$& $69/2^-$& 7.87&1001[E1]& 0.62(7)& E2&0.03(8) \\
727$^a$& $15/2^-$& $11/2^-$& 11.99 \\
732& $71/2^-$& $67/2^-$& 3.65& 1001[E1]& 0.31(2)& E2 \\
734& $53/2^{(-)}$& $51/2^{(+)}$& 2.93& 637[M1]& 0.85(3)& E1& 0.42(4) \\
761& $53/2^+$& $49/2^+$& 76.77& 913[E2]& 0.89(3)& E2& 0.01(2) \\
783$^a$& $(19/2)^-$& $15/2^-$& 3.86 \\
796& $67/2^-$& $67/2^-$& 9.68&  913[E2]& 1.11(8)&&-0.03(2) \\
918$^a$& $23/2^-$& $19/2^-$& 8.26 \\
1001& $63/2^-$& $61/2^+$& 51.79& 913[E2]& 1.24(7)&$0.19_{-0.04}^{+0.05}$&0.05(2) \\
1042& $49/2^+$& $45/2^+$& 77.27& 913[E2]& 0.99(3)& E2&0.02(3) \\
1126& $69/2^-$& $67/2^-$& 9.02& 1001[E1]& 1.02(4)& 0.17&-0.11(1) \\
1185& $69/2^-$& $67/2^-$& 1.89 \\
1325& $69/2^-$& $65/2^-$& 4.60& 1001[E1]& 0.56(9)& E2&0.02(8) \\
1476&$67/2^-$& $61/2^+$& 5.74 \\
1528& $71/2^-$& $67/2^-$& 6.09& 1001[E1]& 0.42(7)& E2 \\
\hline
\hline
\end{tabular}

\end{table*}

\subsection{Lifetime measurement}

\begin{figure}\vspace{2.5cm}

\includegraphics[width=\columnwidth,angle=0]{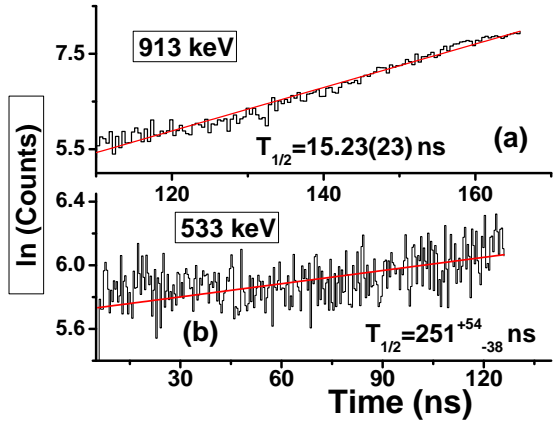}
\vspace{-2.5cm}
\caption{\label{913-533}  (Color  Online)  Determination of (a)(top)
prompt resolution of the TAC spectrum by fitting the decay  curve
of  prompt  913  keV transition decaying from 3685 keV level, and
(b) (bottom) the lifetime of the isomeric level at 2772 keV from  
the RF-$\gamma$ TAC spectra gated by 533 keV transition. The 
logarithmic (ln) values of counts  (to the base e) are plotted 
against time (ns) for ease of fitting. }

\end{figure}

 In  Fig.\ref{913-533},  RF-$\gamma$  TACs correspond to, (i) 913
keV, a prompt gamma which decays from a state above the isomer at
2772 keV and (ii) the 533 keV gamma,  which  is  emitted  from  a
state just below the isomer. The decay curves clearly distinguish
the two (Fig.\ref{913-533}).

The  lifetimes of the isomers have been determined by comparing a
sequence of gamma gated TAC spectra and fitting  them  with  one,
two  or  three component exponential decay curves. The prompt RF-
TAC  spectrum  generated  with  gate  on   913   keV   transition
corresponds  to a resolution of 15.23 $\pm$ 0.23 ns. For 533 keV,
decaying from the isomer through 36 keV, the half-life comes  out
to be $251^{+54}_{-38}$ ns (Fig. \ref{913-533}).

A few other isomers, apart from this long lived isomer, have been
reported  in  the  earlier  work \cite{153hon,153ho}. It has been
discussed already that the relative intensities of the transitions
in the coincidence  spectra  gated  by  a  transition  below  the
isomers   have   been   utilized   to  reconfirm  their  presence
qualitatively.

For  confirming  these isomers further \cite{153ho,153hon} and to
search for new isomers, another technique has  been  adopted.  It
has been mentioned earlier that two time-gated symmetric matrices
with  200 ns and 800 ns gates in the $\gamma-\gamma$ time spectra
have  been  generated.  In  the  Table  \ref{rftact},  ratios  of
intensities  of  different  $\gamma$s  ($I_{\gamma}$)  in 913 keV
gated spectra from 200ns and 800ns time matrices  are  tabulated.
It  is  demonstrated that for the $\gamma$'s below the long-lived
isomer ($T_{1/2} > 200ns$), this  ratio  is  $\approx$  0.66-0.78
indicating  their  increased  yield in the spectra generated from
800ns matrix. On the other hand for  prompt  gammas,  this  ratio
ranges  from  0.94-0.98.  However,  at several parts of the level
scheme, gamma rays of similar energies exist above and below  the
$\simeq$  200 ns isomer. For example, the intensity ratios of 363
and 631 keV gammas can not used  for  validating  isomers  as  at
least  two gamma rays of similar energies are emitted from levels
above and below the long-lived isomer. Systematic  comparison  of
these  ratios  also  indicates  existence  of a relatively longer
lived isomer ($> $300 ps as  reported  in  \cite{153ho}  at  9074
keV).  For  transitions  emitted  from  levels  above 9074 keV in
coincidence with 140 keV gamma decaying from 9074 keV state,  the
ratios  lie  between 0.81-0.93 (Table \ref{rftact}). A comparison
of these ratios with those corresponding to the  isomer  at  2772
keV indicates that the lifetime of the isomer at 9074 keV will be
around   50   ns,  instead  of  $\simeq  300ps$  as  reported  in
\cite{153ho}.

\begin{table}
\caption{\label{rftact}  Ratio of intensity ($I_{\gamma}$) in 913
keV gated spectra from 200ns and 800ns time matrices for $\gamma$
- ray transitions in $^{153}$Ho. Gammas below and above the  long
-lived  isomer  have been indicated as $L$ and $U$, respectively.
The 195 keV emitted from the 3 ns isomer at 7598 keV is indicated
by $I$. The 140 keV gamma from the new isomer has been marked  as
$N$.  $U140$  indicates gammas emitted from states above 9074 keV
state in coincidence with 140keV gamma decaying from this state.}

\begin{tabular} {ccccc}
\hline
\hline
$E_{\gamma}$(keV)&                             \multispan{2}\hfil
Intensity($I_{\gamma}$)\hfil&        \hfil        Ratio        of
$I_{\gamma}$s&Comment \\

&&& (200ns/800ns) \\
& \multispan{2}\hrulefill& \hrulefill \\
& 200ns& 800ns \\
\hline
\hline
378& 8566& 10352& 0.83&L+U\\
701& 6946& 7564& 0.92 &U140\\
711& 5074& 5973&0.85 &SU140 \\
343& 5163& 5541& 0.93&U140  \\
389& 4045& 4609& 0.88 &U140 \\
734& 1986& 2260& 0.88 &732 at U140\\
1528& 1693& 1941& 0.87 &U140 \\
796& 6005& 6453& 0.93 &U140\\
1185& 1392& 1716& 0.81  &U140\\
1126& 5578& 6230& 0.89  &U140\\
1326& 2614& 2771& 0.94  &U140 but bypass 140\\
1476& 1885& 2240& 0.84  &U140\\
140& 9164& 10457& 0.87 &N \\
1001& 31392& 32872& 0.95&U \\
335& 24849& 26834&0.93 &U \\
195& 36258& 37697& 0.96&I\\
497& 1972& 2210& 0.89&U \\
305& 5512& 5629&  0.98&U\\
442& 8671& 9848& 0.88&U+U140 \\
637& 6597& 7019& 0.94 &U\\
466& 38900& 41260& 0.94&U\\
761& 41609& 43533&  0.95 &U\\
1042& 46346& 49211& 0.94 &U\\
455& 48853& 52600&  0.93&U\\
363& 56562& 62031&  0.91&U+L \\
631& 56652& 62900& 0.90&L+U \\
475& 1377& 1964& 0.70&L \\
533& 10269& 13303& 0.77&L \\
557& 15669& 21518& 0.73 &L\\
515& 3988& 5409& 0.74 &L\\
576& 11913& 17014& 0.70&L \\
439& 9268& 11931&0.78&L \\
424& 2394& 3152&0.76&L\\
233& 3001& 4509&  0.66 &L\\
252& 1706& 7043&  0.72 &L\\
918& 2449& 3402& 0.72&L\\
666& 3017& 4246& 0.71 &L\\
727& 4109& 5426&  0.76 &L\\
\hline
\hline
\end{tabular}
\end{table}
\begin{figure}\vspace{2.5cm}

\centering
  \begin{tabular}{c}
\includegraphics[width=.5\textwidth, angle=0]{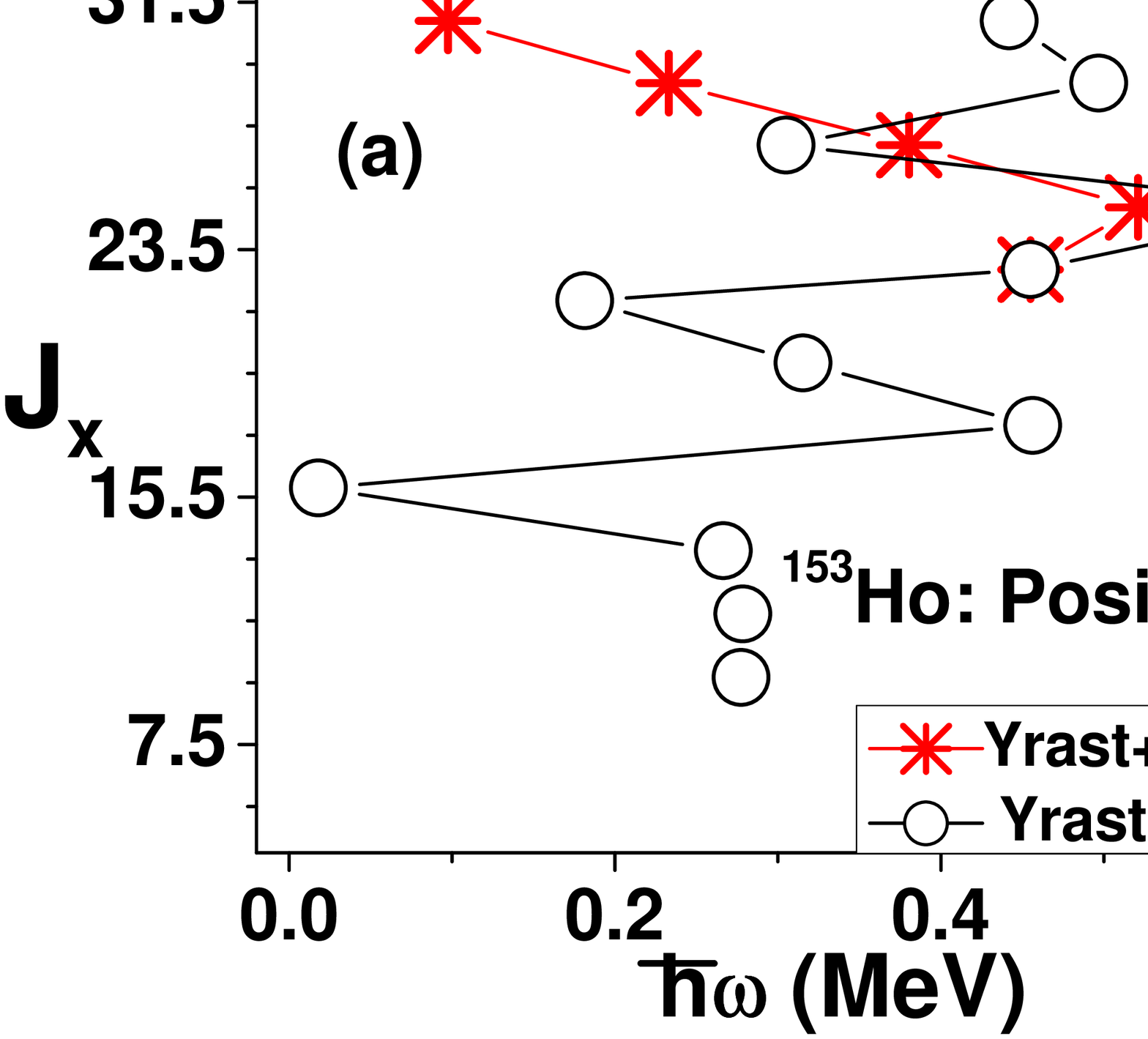} \\
   \includegraphics[width=.5\textwidth, angle=0]{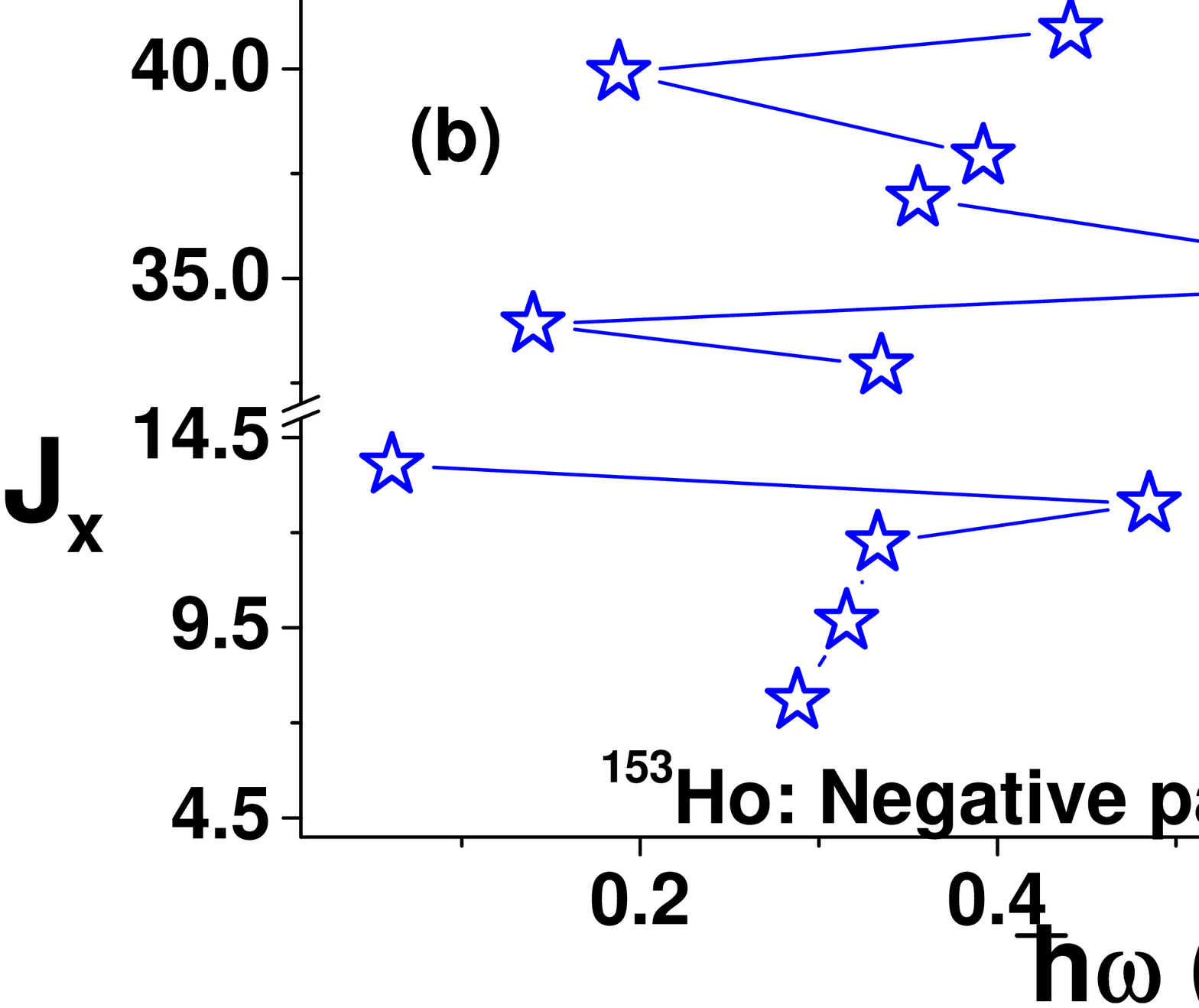}
\end{tabular}
 \vspace{-2.5cm}
\caption{  \label{align}  (Color Online) The alignment plots for
(a) positive  and  (b)  negative  parity  states  in  $^{153}$Ho}
\end{figure}

\section{Discussions and THEORETICAL CALCULATIONS}

The structure of this nucleus poses a few interesting features as
discussed below.

\begin {itemize}

\item{}  Several  isomers have been reported earlier \cite{153ho}
in the level scheme. They are similar to the yrast traps observed
frequently in the nuclei in this mass region. The study of  these
isomers  will  be useful to follow the evolution of the structure
of this nucleus with increasing excitation energy and spin.

\item{}  The  negative  parity states are distributed unevenly in
the scheme. After 27/2$^-$, the next negative parity  states  are
observed  at  spin  53/2$(^-)$  and  then at spin 63/2$^-$. These
large gaps in spins for negative parity  states  are  interesting
and need special attention.

\item{} For the positive parity 49/2$^+$ and 53/2$^+$ levels, the
preferred deexcitation path includes the non-yrast 49/2$^+_2$ and
53/2$^+_2$,  which  contradicts  normal  expectation.  This needs
special attention and theoretical interpretation.

\item{}  The  level  scheme  apparently is not regular and it has
been emphasized  earlier  \cite{153ho}  that  the  non-collective
character  of  the  motion manifests itself in this irregularity.
However, in neighboring Dy isotopes, non-yrast collective states
have been found to coexist with non-collective yrast  states.  So
whether  Ho also follows this trend or not needs to be understood
phenomenologically as well as utilizing reliable theory.

\end {itemize}

The  alignment plot of the positive and negative parity bands are
shown in Fig. \ref{align}. Here $J_x$ has been  calculated  using
the  expression $J_x= \sqrt{(J(J+1) -K^2)}$. Positive parity band
has been plotted taking K=7/2, whereas for negative parity  band,
K=5/2 has been considered. These figures are useful to reveal the
structural features.

\begin {itemize}

\item{}For  the  positive parity states, up to 45/2$^+$ state the
excitation  pattern  consists  of  several   twists   and   turns
indicating  several  changes  in structure. Interestingly most of
the bends are associated with isomers. The longest life isomer is
at the first bend ($J_x \simeq 15.5)$.

\item{} At 45/2$^+$ , the excitation scheme bifurcates, the yrast
branch is irregular indicating single particle alignment, whereas
the non-yrast one is more regular and probably signify collective
structure.

\item{}  The  negative  parity  states  show  smooth  increase in
alignment till 23/2$^-$. Beyond that the alignment is  irregular,
indicating single particle mode of angular momentum generation.

\item{}  A  large  discontinuity  in alignment is observed beyond
27/2$^-$ state.  The  next  negative  parity  state  has  a  spin
53/2$(^-)$, about 13 units of angular momentum is gained.

\end {itemize}

The  structure  of  this  nucleus  has been studied theoretically
using two models. The issue of searching for  energetically  most
stable  shape  at  each  angular  momentum has been understood by
calculating total Routhian  surfaces  \cite  {trs}  at  different
angular  momenta.  Moreover,  a  version  of Particle Rotor Model
\cite{prm1,prm2} has been  utilized  to  calculate  energies  and
transition probabilities of specific states.

\begin{figure*}
\centering
  \begin{tabular}{@{}cc@{}}
\vspace*{-2cm}\hspace*{-2cm}\includegraphics[width=.5\textwidth, height=0.5\textwidth, angle=-90]{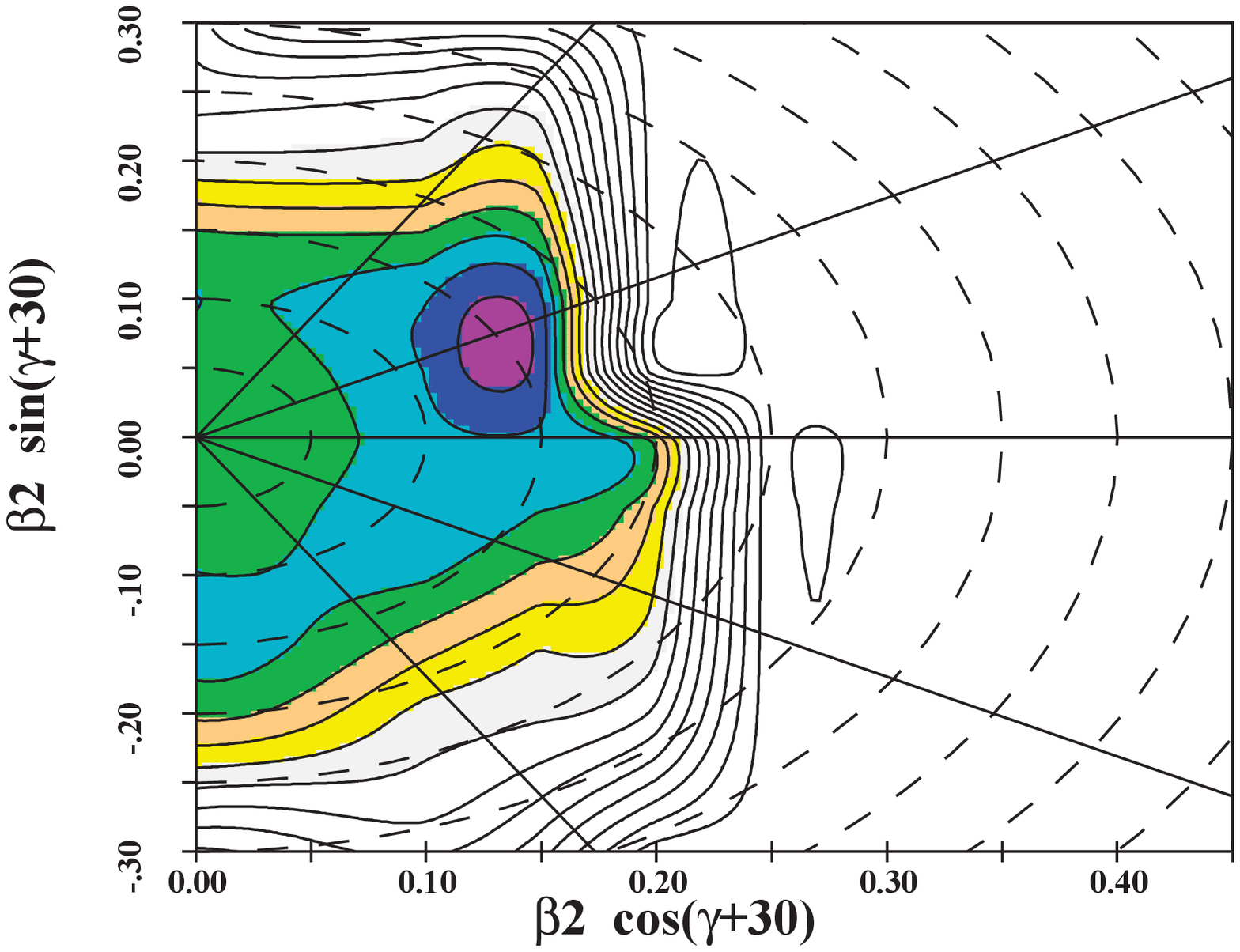}  &
    \hspace*{-.5cm}\includegraphics[width=.5\textwidth, height=0.5\textwidth, angle=-90]  {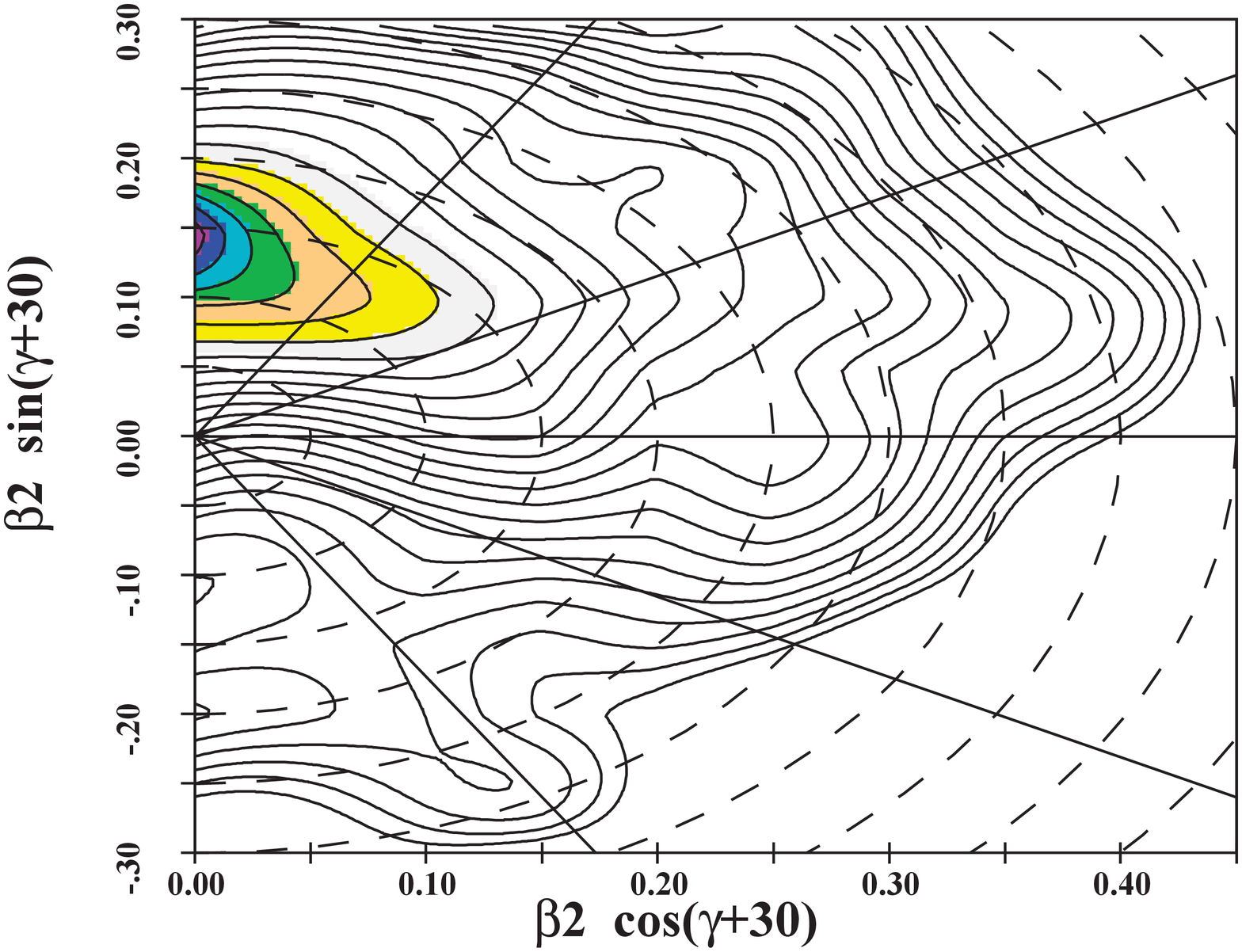}\\
\hspace*{4cm}{\bf \huge(a)}&\hspace*{2cm}{\bf\huge(b)}\\
\\\\\vspace*{-2cm}
\hspace*{-2cm}\includegraphics[width=.5\textwidth, height=0.5\textwidth, angle=-90]{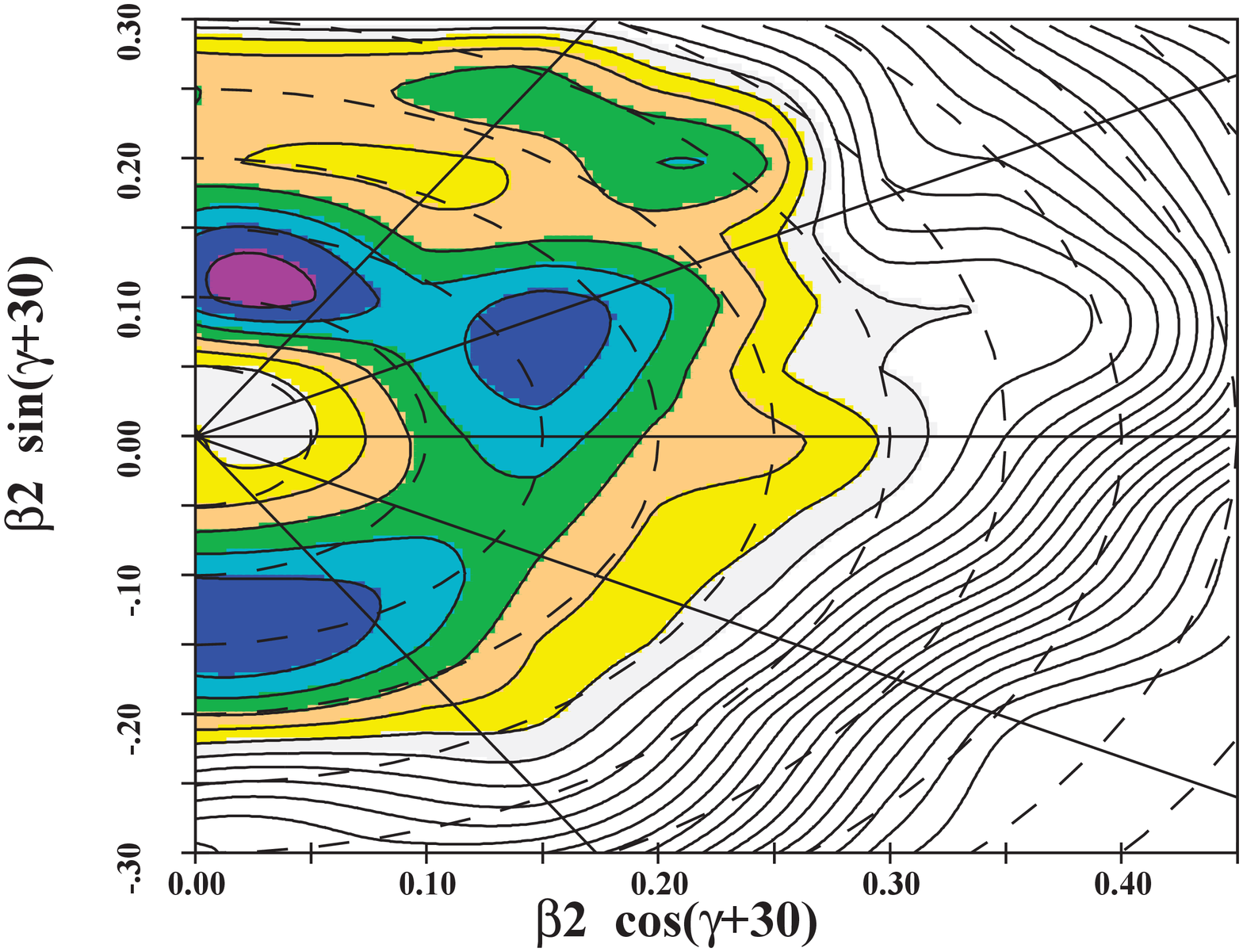}&
   \hspace*{-1cm} \includegraphics[width=.5\textwidth, height=0.5\textwidth, angle=-90]{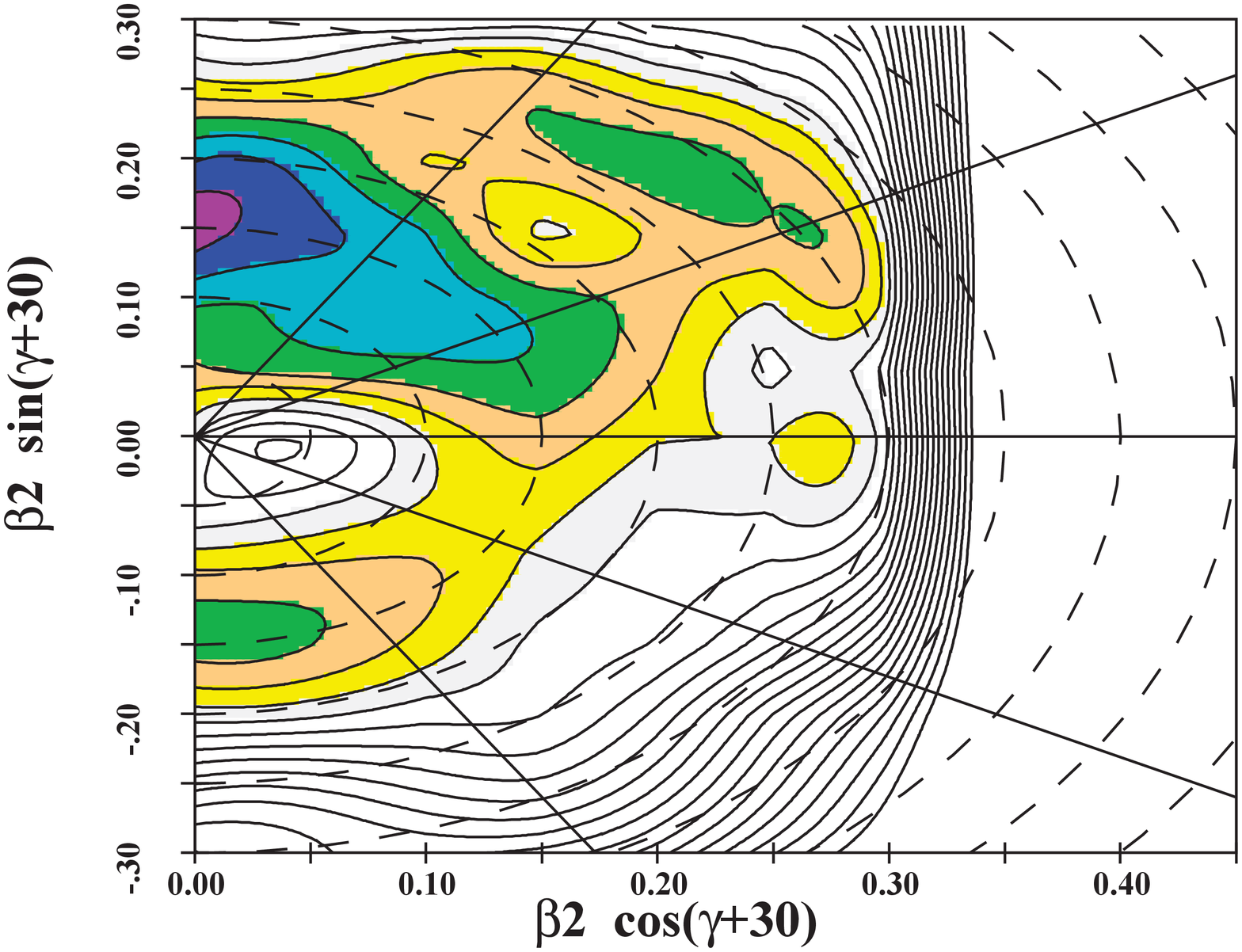}
\\\hspace*{ 4cm}{\bf \huge(c)}&\hspace*{ 4cm}{\bf\huge(d)}\\
\end{tabular}
  \vspace*{1cm}\caption{
\label{trs1}(Color  Online)  TRS  plots  for  the positive parity
states. In the top row the plots correspond to (a)$\hbar \omega=$0.2
(left) and (b) 0.40 (right) MeV  (with  $J  \simeq  2.5$  and  37.5,
respectively).  In the bottom row, they are calculated for (c) $\hbar
\omega=$ 0.25 and (d) 0.28  MeV,  (with  $J  \simeq  9.5$  and  21.5,
respectively).  Shape coexistence at $\omega$=0.25 MeV is clearly
indicated with yrast oblate  and  non-yrast  prolate  minimum.  }
\end{figure*}

\subsection{TRS Calculations}

\begin{figure*}
\centering
  \begin{tabular}{cc}

\vspace*{-2cm}
\hspace*{-2cm}\includegraphics[width=.5\textwidth, height=0.55\textwidth,angle=-90]{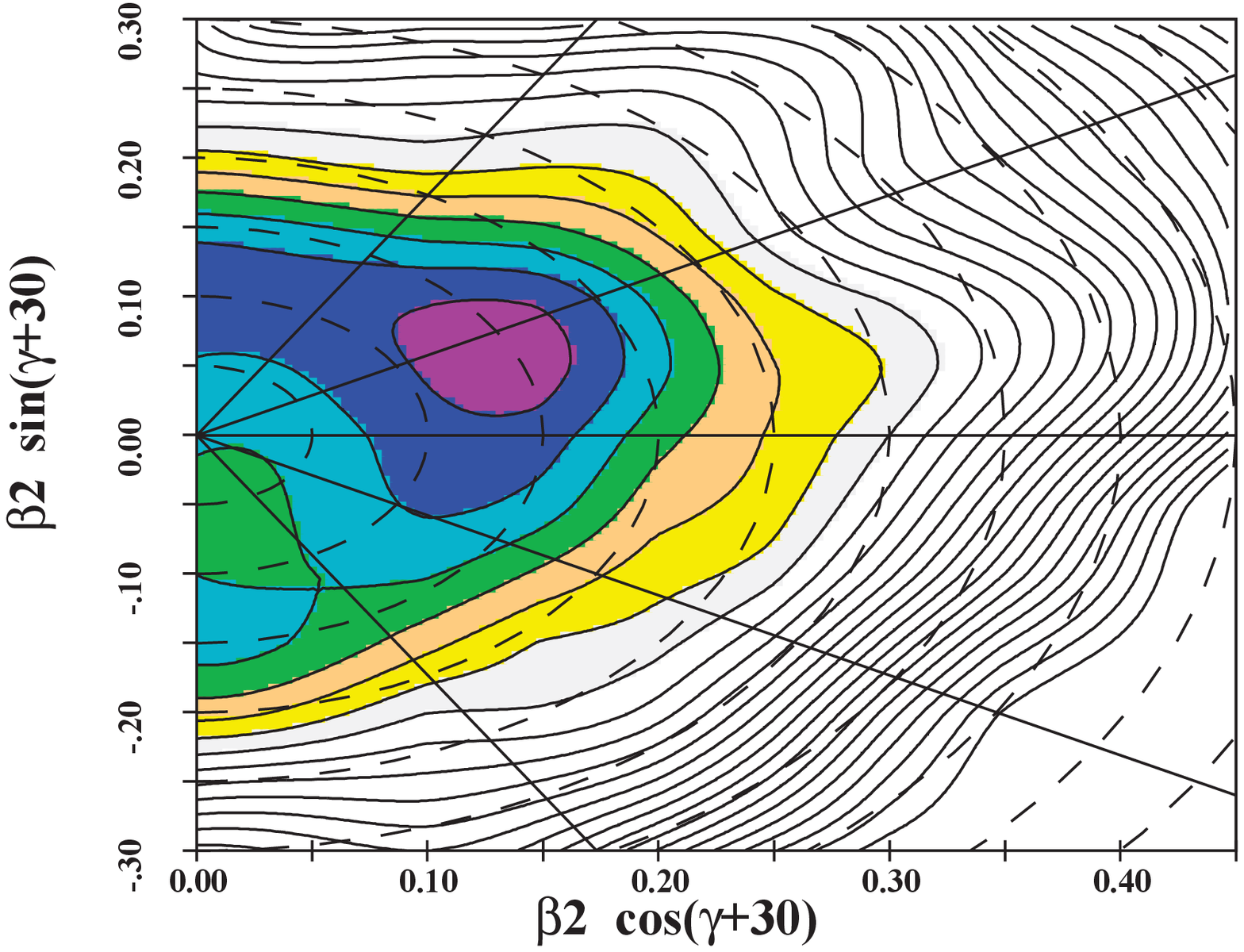} &
    \hspace*{-1cm}\includegraphics[width=.5\textwidth, height=0.55\textwidth,angle=-90]{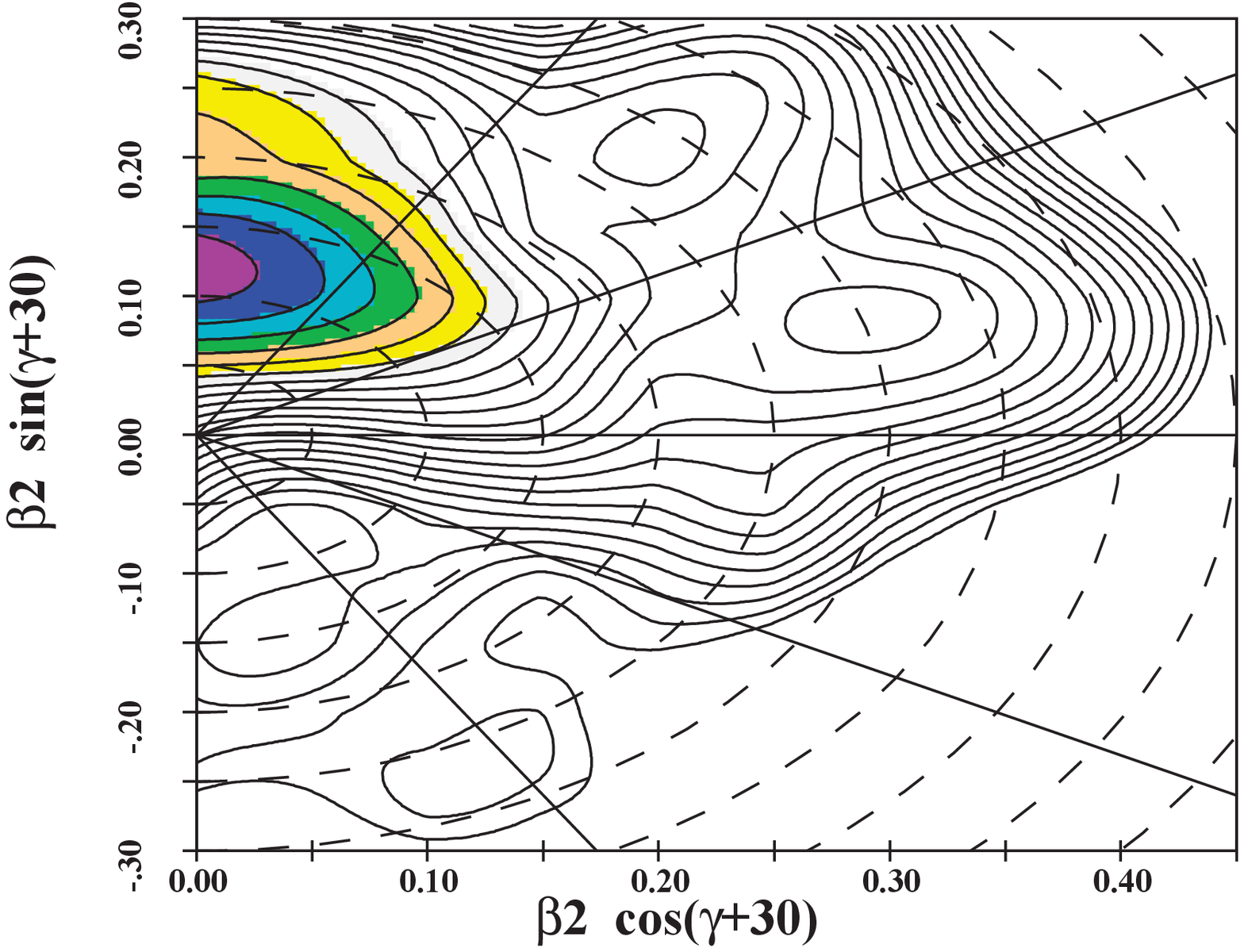}\\
\hspace*{4cm}{\bf \huge(a)}&\hspace*{2cm}{\bf\huge(b)}\\
\\\\
\vspace*{-2cm}
\hspace*{-2cm}\includegraphics[width=.5\textwidth, height=0.55\textwidth,angle=-90]{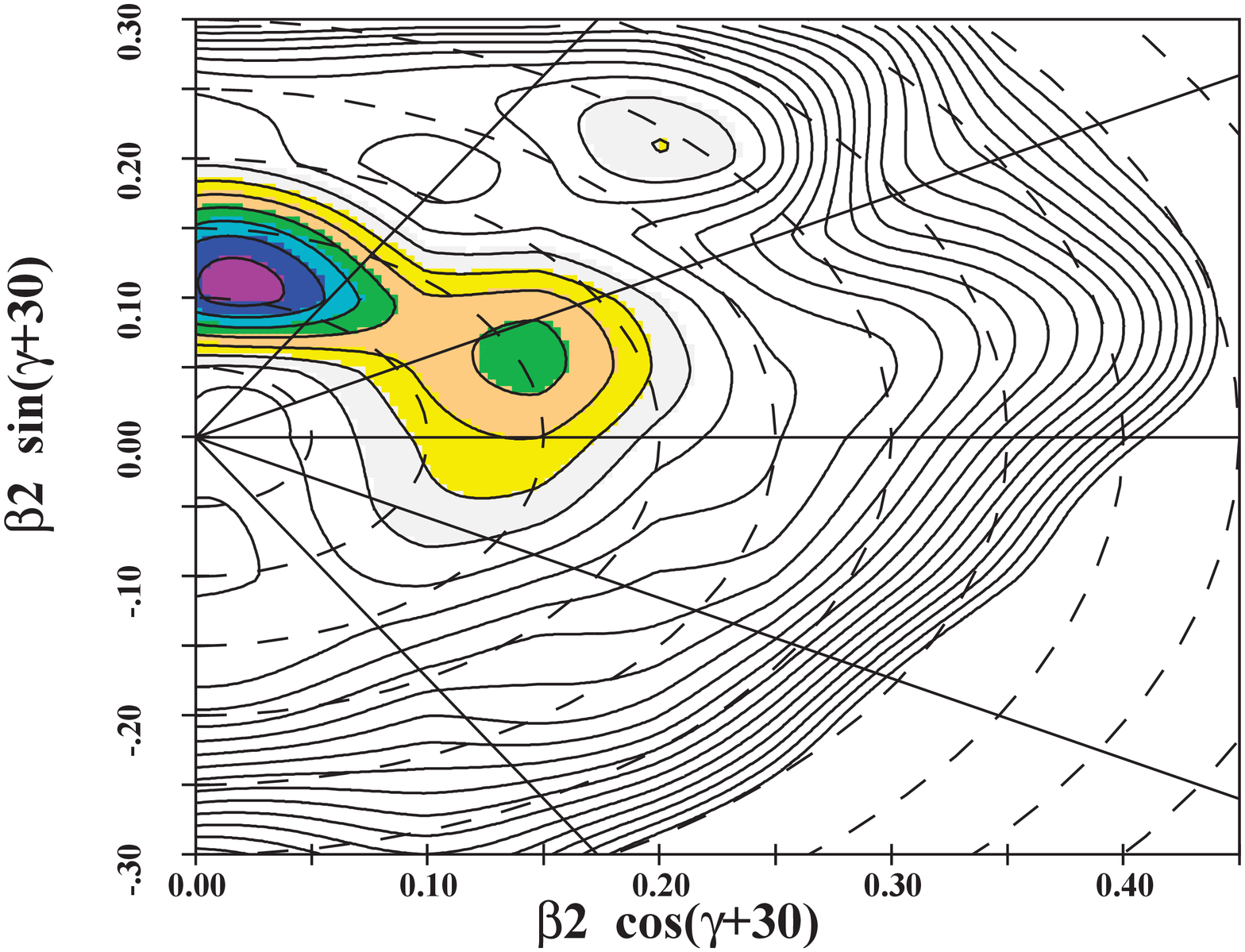} &
    \hspace*{-1cm}\includegraphics[width=.5\textwidth, height=0.55\textwidth,angle=-90]{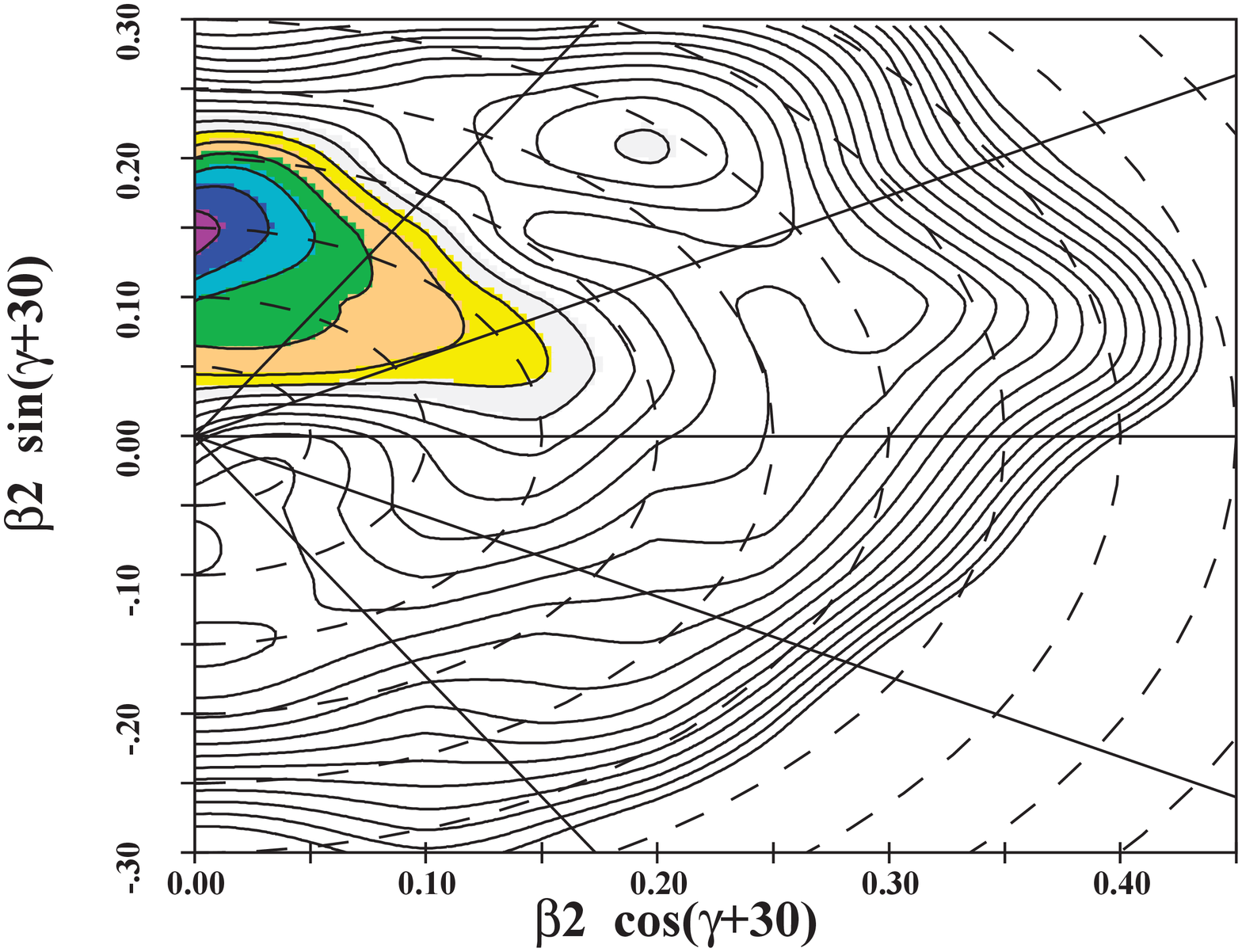}\\
\vspace*{-4cm}
\hspace*{ 4cm}{\bf \huge(c)}&\hspace*{ 4cm}{\bf\huge(d)}\\
\end{tabular}
\vspace*{5cm}
 \caption{  \label{trs2}(Color Online) TRS plots for the negative
parity states. The  plots  correspond  to  (i)  Top  Row: (a) $\hbar
\omega=$0.2  (left)  and (b) 0.45 MeV(right),(with $J \simeq$ 6.5 and
32.5, respectively). (ii) Bottom row: (c) $\hbar \omega= $0.25  (left)
and  (d)  0.28   (right)   MeV,(with   $J   \simeq$  12.5  and  24.5,
respectively)} \end{figure*}

 The Total Routhian Surface (TRS) calculations \cite {trs} with a
Woods   -   Saxon   potential  and  monopole  pairing  predict  a
near-prolate  deformation  of  $\beta_2  \simeq  0.20$  for  this
nucleus.   As  observed  in  Figs.\ref{trs1},\ref{trs2},  at  low
frequencies with $\hbar \omega =$  0.20  -  0.24  MeV,  both  the
negative  and  positive parity states favor prolate deformation.
However, at $\hbar \omega= 0.25$ MeV, shown for  positive  parity
states  (Fig. \ref{trs1}), there is a sudden transition to oblate
shape as the most favored one. At higher rotational frequencies,
the oblate shape remains  energetically  favored  for  both  the
parities  (Figs.  \ref{trs1},\ref{trs2}).  This can be associated
with the rotation alignment of a  pair  of  $i_{13/2}$  neutrons,
increasing  the  angular momentum from J $\simeq$ 9 to 21, over a
small  rotational  frequency  range  of   0.25   to   0.28   MeV,
contributing  12  units  of  angular momentum. These observations
have been important while extending our studies using PRM.

The  presence  of  several positive parity single particle states
near  the  Fermi  level  smoothens  shape  transition  and  shape
coexistence  is observed (Fig. \ref{trs1}). However, for negative
parity, single  particle  orbitals  arising  only  from  intruder
$1h_{11/2}$  state  are available. The shape transition therefore
is drastic here and it also possibly leads to large  gap  in  the
available  negative  parity  states  in  the spectrum in the band
crossing  region  as  predicted   by   TRS   calculations   (Fig.
\ref{trs2}).  There  may  be  other negative parity states in the
spin gap which has almost no overlap with  higher  spin  negative
parity  states.  Possibly  they  are not populated in the present
heavy-ion fusion evaporation reaction for this reason.

\subsection{PRM Calculations}

To  calculate  specific  energies and transition probabilities of
different levels, PRM calculations have been done. The particular
version of the model \cite{prm1,prm2} is discussed below.

\subsubsection{\bf{Formalism}}

The  model  is  based  on  the  assumption that the nucleus under
consideration is axially symmetric. In this model, the motion  of
an  unpaired quasiparticle in a Nilsson deformed orbit is coupled
to  the  rotational  motion  of   the   core   through   Coriolis
interaction.  We  have used a version of the PRM \cite{prm1,prm2}
in which the experimental core energies can be  fed  directly  as
input parameters.

The Hamiltonian of the odd-A system can be written as

\begin{equation}
{H} = {H}_{qp}^{0} + {c}{\bf{R.j}} + {E}_{c}{(R)}
\end{equation}

The  first term is the Hamiltonian of a single quasiparticle. The
quasiparticle (quasiproton, in the present case) is assumed to be
moving in  an  axially  symmetric  Nilsson  potential  under  the
influence of BCS pairing. The pairing gap and the Fermi level are
represented by $\Delta$ and $\lambda$ respectively.

The  total  Hamiltonian  is  then diagonalize, giving the energy
eigenvalues and the wave functions of the final states $|JM>$  in
terms of the Coriolis mixing amplitudes {$ f_{JK}$} and the basis
states $|JMK>$:

\begin{equation}
|JM> = \sum_{K}f_{JK}|JMK>
\end{equation}

In  the present version of the model \cite{prm2}, to identify the
rotational composition of the final state  $|JM>$,  these  states
are expanded in terms of states with sharp {\it R} and {\it j}:

\begin{equation}
 |JM> = \sum_{jR}\sum_{K}f_{JK}\alpha_{jR}^{(K)}|JMjR>,
\end{equation}

where
\begin{equation}
\alpha^{(K)}_{jR} = \sqrt{2}
\left [ {\begin{array}{ccc}
J& j& R \\
K& -K& 0 \\
\end{array}}\right ]
\end{equation}

So  to  calculate  a  state  with total angular momentum {\it J},
where the single-particle angular momentum involved  is  {\it  j}
(say),  the  experimental core energies required will be given by
the following range of {\it R} values:

\begin{equation}
{R}_{max} = {J} + {j},
\end{equation}

\begin{equation}
{R}_{min} = {J} - {j}.
\end{equation}

\begin{figure}\vspace{2.5cm}

\includegraphics[width=\columnwidth,angle=0]{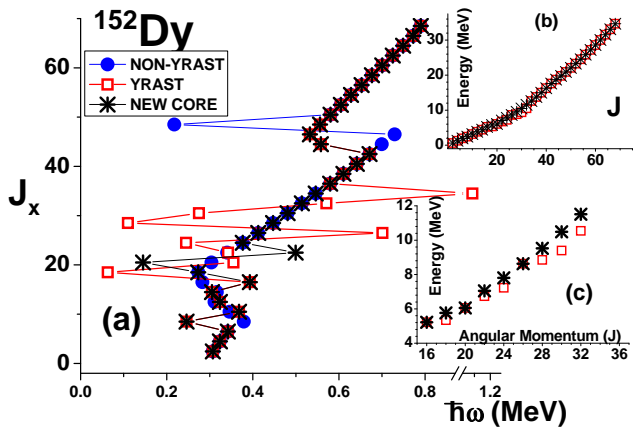}
 \vspace{-2.5cm}
\caption{\label{core}  (Color  Online)  (a)  The  alignment plot
shows the yrast, nonyrast states of $^{152}$Dy. The  core  states
($new  core$)  selected  for  our  calculations  are indicated by
stars. For all cases K=0 value has been considered for  alignment
calculation.  (b)  Experimental energy spectra of $^{152}$Dy as a
function of  angular  momentum  (J).  (c)  The  figure  shows  an
expanded  view  of  the  region  where  vibration  and rotational
features compete and coexist. } \end{figure}

\subsubsection{\bf{Parameter choice}}

There  are  several  parameters involved in the PRM calculations.
The single-particle Nilsson  parameters  $\mu$  and  $\kappa$  (=
0.5920  and  0.065,  respectively),  have  been  deduced from the
expression provided by  Nilsson  {\it  et  al.}  \cite{nil}.  The
deformation  parameter  for  the  odd  nucleus is chosen from the
systematics  of  the  experimentally  deduced   values   in   the
neighboring  even isotopes \cite{raman}, which agrees quite well
with  the  deformation  ($\beta_2$)   obtained   from   the   TRS
calculations (Figs. \ref{trs1}, \ref{trs2}). In our calculations,
$\delta  (= 0.95 \beta_2)= \pm 0.14$. The pairing gap $\Delta$ is
deduced  from   the   experimental   odd-even   mass   difference
($\Delta_{o-e}$)  calculated  from  mass data \cite{audi}. In our
case, the experimental odd-even mass difference $\Delta_{o-e}$  =
1.5  MeV.  For  positive  (negative)  parity  states, all Nilsson
single quasi-particle states from N=4 (N=5) have been considered.
We have selected the Fermi level $\lambda$ in such a way that  it
remains  consistent  with the observed ground state spin. We have
calculated the energy spectra with the Fermi level at $\lambda$ =
43.8 MeV, for both  positive  and  negative  parity  states.  For
positive  parity,  it lies close to $3/2^+[411]$ and $3/2^+[422]$
orbitals originating primarily  from  2d$_{5/2}$  and  $1g_{7/2}$
spherical   states,   respectively.   The  Fermi  level  is  near
$5/2^-[532]$ Nilsson orbital  originating  from  $1h_{11/2}$  for
negative parity states.

Here,  the  only  adjustable  parameter  is  Coriolis attenuation
coefficient $\alpha$. For $^{153}$Ho nucleus, $\alpha$  is  taken
to  be  0.90  and  0.99  for  positive and negative parity bands,
respectively. While  calculating  transition  probabilities,  for
electric  transitions  effective charge $e_{eff}$ = 1.0 have been
used. Intrinsic quadrupole moment ($Q_0$) is necessary for  these
calculations  has  been  provided  from  experimental  quadrupole
moment of $^{152}$Dy \cite{raman}. For magnetic transitions, free
values of $g_{l}$ and $g_{s}$ have been used.

\begin{figure}\vspace{2.5cm}

\centering
 \includegraphics[width=\columnwidth]{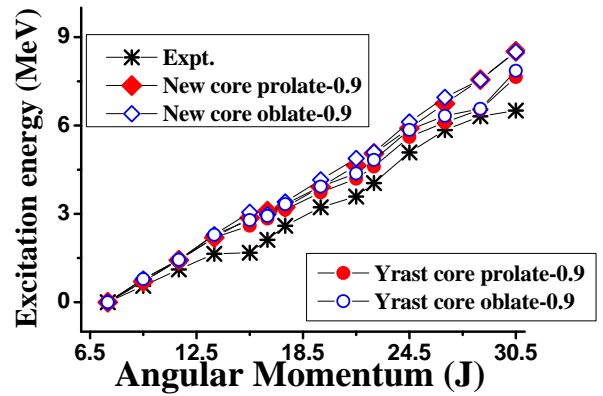}
\vspace{-2.5cm}
\caption{
\label{nonorm}(Color Online) The energy vs angular momentum plots
for  the  positive parity states for different shape options with
yrast and new cores.} \end{figure}

\subsubsection{Choice of Core Energies}
 The  core  energies are usually taken as the excitation energies
of the yrast band of the neighboring even isotope. The $^{153}$Ho
nucleus is represented as an odd proton coupled to  a  $^{152}$Dy
core.  In  the  present  work,  the  level  energies of the yrast
spectra of $^{152}$Dy are used as core energies.

While  choosing the yrast spectra of Dy shown in Fig. \ref{core},
it is  found  that  it  has  an  interesting  feature,  the  core
possesses   rotational   as   well   as   vibrational   character
\cite{dy152b}.  Although  at   certain   angular   momentum   the
vibrational core states are yrast, the states populated via heavy
ion  fusion  evaporation  reaction  also  deexcites  via nonyrast
states which are members of the rotational band \cite{nndc}.  Two
parallel  branches  of  excitation  scheme  are  observed, mildly
collective and non-collective states are prevalent at low  spins,
at  higher  spins the coexisting non yrast states show collective
behavior \cite{dy152b}. This is similar to the parallel branches
observed  in  $^{153}  Ho$  (Fg.  \ref{153ho-level})  for   spins
49/2$^+$ and 53/2$^+$ (Fig. \ref{align}).

\begin{figure*}\vspace{2.5cm}

\centering
  \begin{tabular}{@{}cc@{}}
\includegraphics[width=.5\textwidth, angle=0]{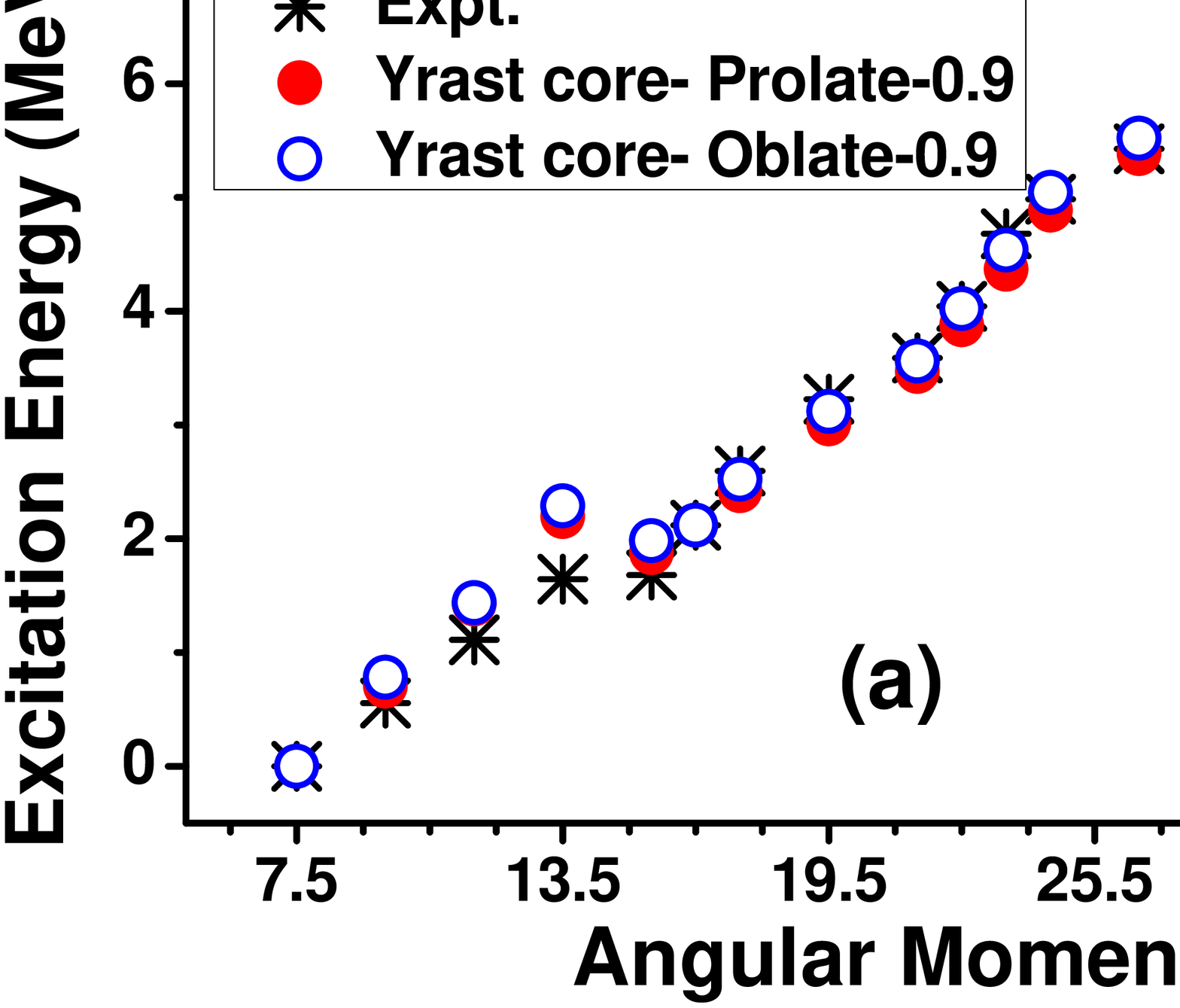} &
    \includegraphics[width=.5\textwidth, angle=0]{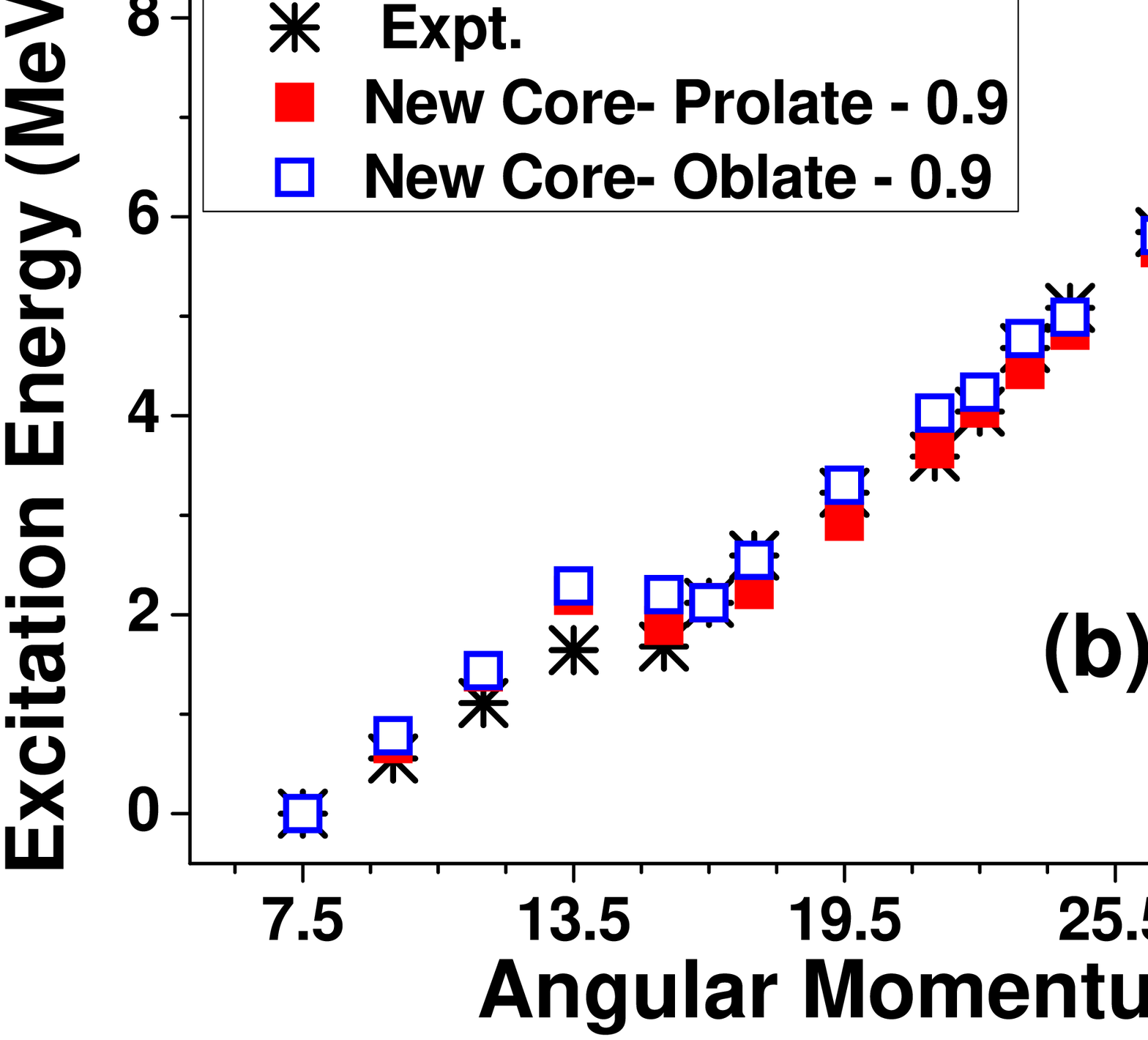}
\end{tabular}
 \vspace{-2.5cm}
\caption{
\label{norm}  (Color Online) The energy vs angular momentum plots
for the positive parity states for different  shape  options.  a)
Results   from   calculations   with  yrast  core  compared  with
experimental yrast states. (b)  Results  from  calculations  with
$new  core$  compared with experimental states including nonyrast
$49/2^+_2$ and $53/2^+_2$.} \end{figure*}

The  TRS  calculations have been useful in this regard suggesting
shape coexistence at certain  angular  frequency  in  $^{153}$Ho.
These results suggested that the alignment of $i_{13/2}$ neutrons
are  responsible  for  shape  changes in $^{153}$Ho. So the shape
coexistence  observed  in   even-   proton   nucleus   $^{152}$Dy
\cite{dy152b}  caused  by neutron alignment is also manifested in
neighboring odd-proton $^{153}$Ho. Usually, the PRM calculations
are not supposed to reproduce alignment effects. However, it  has
been  demonstrated  earlier  \cite{prm2} that, if the backbending
(bb) is observed in the core for neutron (proton) alignment,  for
neighboring  odd-  proton  (neutron) nucleus, the backbending is
reproduced using the present version of PRM  provided  the  input
core  states  included the backbending. This shape transition can
be reproduced theoretically by this version of PRM, provided  the
proper core states are included in the calculation.

Therefore,   spectra   for   $^{153}$Ho   have   been  calculated
considering only yrast states of the core (indicated as  $yrast$)
and  both the yrast states and non yrast rotational states of the
core, indicated as $new core$(Fig. \ref{core}). Calculations were
performed with prolate as well  as  oblate  deformation  for  the
single    particle   Nilsson   potential   (Figs.   \ref{nonorm},
\ref{norm}). In Table \ref{trans}, Table  \ref{bran1}  and  Table
\ref{mix},  the  results  for B(E2)s, branching ratios and mixing
coefficients, respectively of several gamma transitions have been
compared with experimental values.

\begin{table*}
\caption{\label{trans} Comparison of experimental and theoretical
reduced  transition  probabilities  (B(E2)s  in ($e^2 b^2$)). For
experimental quantities, corresponding errors are included within
parentheses.   The   average   core   angular   momentum    ($R$)
corresponding to each state has also been tabulated for different
options.  }  \begin{tabular}  {c c c c|c c| c|c|c|cc| c| c |c|c }
\hline

&&&&\multispan{2}\hfil Expt.\cite{nndc,153ho}\hfil&\multispan{9} \hfil Theory \hfil \\
&&&& &&\multispan{4}\hfil Yrast core\hfil&&\multispan{4}\hfil  New core\hfil\\
$E_x$&$ J_i$& $J_f$& $E_\gamma$&$T_{1/2}$&B(E2)\footnote{extracted from experimental half-lives.} &\multispan{2}\hfil
Prolate\hfil&\multispan{2}\hfil Oblate\hfil&&\multispan{2}\hfil
 Prolate\hfil&\multispan{2}\hfil  Oblate\hfil\\
(keV)&&&(keV)&($ns$)&($e^2 b^2$)&B(E2)&$R_i(R_f)$&B(E2)&$R_i(R_f)$&& B(E2)&$R_i(R_f)$& B(E2)&$R_i(R_f)$\\
\hline
2772 &$31/2^+_1$& $27/2^+_1$& 36& $\footnote{From present work}251^{+54}_{-38}$&0.015 (3)& 0.0003&16.1(12.7)& 0.0002
&17.0(11.6)&&0.014&15.9(12.6)&0.102
 &14.0(11.6)\\
4679 &$43/2^+_1$& $39/2^+_1$& 363& 0.5 (2)& 0.02 (1)& 0.106&21.9(20.1) &0.007&22.9(20.4)&&0.003&21.9(20.6)&0.029
&22.6(21.1)\\
7598 &$61/2^+_1$& $57/2^+_1$ &195 & 2.95 (15) &0.052(3) &0.038&31.4(28.7) &0.001&30.7(30.5) &&0.109&30.9(28.9)
&0.117&32.5(30.5)\\
9074&$67/2^-_1$& $65/2^-_1$&140\footnote{M1+E2 transition, mixing ratio =0.2}&0.3 (1)& 0.07(2) &0.001 &33(32)
 &0.002&33(32) &&0.001&33(32) &0.0006&33(32)\\
\hline
\end{tabular}

\end{table*}

\subsection{\bf{Positive-parity states}}

We  have shown the results for the positive parity states in Fig.
\ref{nonorm}. It is noted that above  $23/2^+$,  all  states  are
somewhat  over  predicted  with  both  the  cores. For lower spin
states, results for both prolate and oblate structure are  almost
same for both the cores.

However,  we  found  that  only  if the states above the isomeric
state $31/2^+$ are isolated from the set below, and  the  results
are  normalized  to  the energy of $33/2^+$ state, the normalized
results match reasonably well with the  experimental  data  (Fig.
\ref{norm}). This indicates structural differences between states
above and below the isomer. Specific deformations which give best
agreement at different spins have been discussed below in detail.

\begin{table*}
\caption{\label{bran1} Comparison of experimental and theoretical
branching ratios from calculations with $new core$ for 57/2$^+_1$
state  to investigate the reason behind strong decay path through
non-yrast states. Note that in this set,  as  some  of  the  core
states  are  non-yrast,  experimental  non-yrast  53/2  state  is
equivalent to theoretical yrast state. See text for details. }

\begin{tabular} {c c c c c c c }
\hline

$E_x$&$ J_i$& $J_f$& $E_\gamma$& Exp& \multispan{2}\hfil  New core\hfil\\
&&&&&Prolate& Oblate\\
     \hline

7403 &57/2$^+_1$ &53/2$^{+  (theo)}_1$(53/2$^{+ (expt)}_2)$&466 &100&$\simeq$100&$\simeq$ 100\cr
7403 &57/2$^+_1$ &53/2$^{+  (theo)}_2$(53/2$^{+ (expt)}_1$)&885 &Not observed&$ 2.07 \times10^{-5}$&$ 7.13\times 10^{-5}
$\\
\hline
\end{tabular}

\end{table*}

\subsubsection{Excitation  energies  as  a function of spin (with
normalization at $33/2^+$) and transition probabilities}

\begin{itemize}

 \item  {}  For yrast core, the strongly populated $49/2^+_2$ and
$53/2^+_2$ states (the decay out gamma  transition  energies  are
761  and  1042  keV,  respectively) are non yrast positive parity
states (Fig. \ref{153ho-level}). It has been found that the  high
spin  yrast  positive  parity  states  upto $49/2^+_1$ state have
better  agreement  with  oblate  structure  (Fig.   \ref{norm}a).
Although  the  prolate  option  also reproduces the results well.
Theoretical energies for $53/2^+_2$,  $59/2^+_1$  and  $61/2^+_1$
states deviate from experimental values. However, the B(E2) value
extracted  from  the lifetime of $61/2^+_1$ state decaying by 195
keV gamma is reproduced  well  by  the  prolate  option  of  this
calculation (Table \ref{trans}).

\item  {}In  new  - core calculation (Fig. \ref{norm}b), in which
some of the non-yrast core states are taken as input, theoretical
energies for $49/2^+_1$ and $53/2^+_1$ states have been  compared
with  the  experimental  $49/2^+_2$  and  $53/2^+_2$  states. The
agreement  is  reasonably  better  till  39/2$^+$   with   oblate
deformation. For 43/2$^+$, 45/2$^+$ and 47/2$^+$ states, prolate
deformation is favored. Both prolate and oblate deformation give
almost similar  agreement  for  49/2$^+_2$  and  53/2$^+_2$.  The
calculated  energy for $59/2^+_1$ with prolate deformation agrees
quite well to experiment . The  lifetimes  of  the  $31/2^+$  and
$43/2^+$   states   decaying  by  36  keV  and  363  keV  gammas,
respectively, are reproduced  best  by  the  prolate  and  oblate
deformation  (Table \ref{trans}). So the theoretical results also
(Table  \ref{trans})  indicate  that  the  $31/2^+$  state  is  a
long-lived  isomer.  For  $61/2^+_1$  state  decaying  by 195 keV
gamma,  results  (energies  as  well  as  lifetimes)   for   both
deformation   worsens   further  compared  to  yrast  core  (Fig.
\ref{norm}a).

\item  {}  The  contribution  of the different core states in the
composition of the states of interest in $^{153}$Ho are tabulated
in (Table \ref{trans}). The average core angular  momentum  value
gives  an  idea  about  the  most  relevant  core  states  for  a
particular state in the odd-A nucleus. As expected,  unlike  that
for  negative  parity  states (which only originate from intruder
orbital 1$h_{11/2}$ of N=5),  for  positive  parity,  this  value
changes  for different options of deformation. This is due to the
fact that the most important lowest energy positive parity single
quasiparticle Nilsson states originate from all  states  of  N=4,
except, 1$g_{9/2}$.

\item  {}While calculating the branching of the decay of 7403 keV
(57/2$^{+ (expt)}_1$) level in the calculation with new core, the
branching to the 6937 keV (53/2$^{+  (expt)}_2$)  is  100\%  with
extremely  small branch populating 6518 keV (53/2$^{+ (expt)}_1$)
state (Table \ref{bran1}). This justifies the decay of the  yrast
state  at  7403  keV  to  the  non-yrast state 6937 keV (53/2$^{+
(expt)}_2$). \item  {}  For  the  branch  parallel  to  the  main
excitation  scheme, consisting of 49/2$^{(+)}_1$, 47/2$^{(+)}_1$,
51/2$^{(+)}_1$,  53/2$^{(+)}_1$  etc  the  yrast  states  in  the
calculations   with  yrast  core,  reproduce  the  mixing  ratios
reasonably  well  (Table  \ref{mix}).  Interestingly,  the  large
mixing  of  455  keV  gamma decaying from $45/2^+$ state has been
reproduced by the prolate deformation option of  the  yrast  core
calculations.

\end{itemize}

\begin{table}
\caption{\label{mix}  Comparison  of experimental and theoretical
mixing ratios for different transitions in $^{153}$Ho  for  yrast
core. }

\begin{tabular} {c c c c c c c  }
\hline

$E_x$&$ J_i$& $J_f$& $E_\gamma$& Exp& \multispan{2}\hfil Yrast core\hfil\\
&&&&     &Prolate &Oblate\\
\hline

5134& 45/2$^+_1$&$ 43/2^+_1$& 455 &0.33(3)& 0.29& 0.12 \\
5771& 47/2$^{(+)}_1$& 45/2$^+_1$ &637 &0.08$^{+0.12}_{-0.09}$& 1.62& 0.02\\
6076& 49/2$^{(+)}_1$& 47/2$^{(+)}_1$& 305& $0.11^{+0.11}_{-0.05}$ &0.08& 0.01 \\
6573& 51/2$^{(+)}_1$& 49/2$^{(+)}_1$& 497& 0.3(5) &0.07 &0.12\\
\hline
\end{tabular}

\end{table}

\begin{figure}\vspace{2.5cm}

\includegraphics[width=\columnwidth,angle=0]{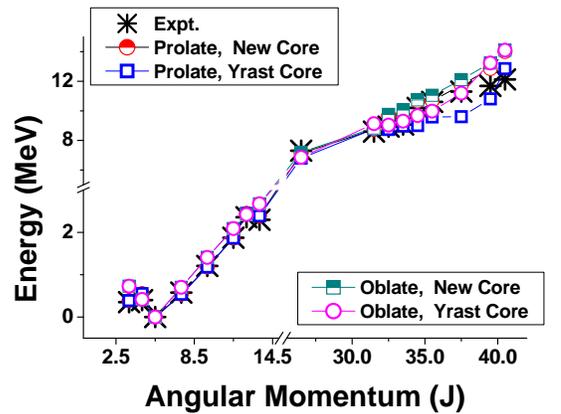}
\vspace{-2.5cm}
\caption{\label{negative}(Color     Online)     Comparison     of
experimental and theoretical energies of negative  parity  states
in $^{153}$Ho}. \end{figure}

\subsection{\bf{Excitation  energies and transition probabilities
of Negative-parity states}}

 For negative parity levels, it is found that for lower spins the
nucleus  is  prolate,  which  changes  to  an oblate structure at
higher  spins  (Fig.  \ref{negative})  as   expected   from   TRS
calculations.  For  the PRM calculations, the choices of yrast or
non-yrast cores do not make much difference in  the  results  for
these  states. Only at the highest spins ($81/2$), prolate option
appears to reproduce  the  energy  better.  For  negative  parity
states,  single  quasiparticle  Nilsson  states  near Fermi level
originate  only  from  intruder  1$h_{11/2}$  (N=5),  leading  to
relatively   less  fluctuations  in  core  angular  momentum  for
different options (Table \ref{trans}). Thereby making the results
less sensitive to the change in the choice of core.

In  the  earlier  work \cite{153ho}, a $\simeq$ 300 ps isomer has
been reported at 9074  keV  state  decaying  by  140  keV  gamma,
corresponding  to a B(E2) value of 0.07 (2)$e^2b^2$ ($\simeq$ 659
e$^2fm^4$)   (Table   \ref{trans})   extracted   utilizing   the
experimental  mixing  ratio determined in the present work (Table
\ref{intense}).  None  of  the  calculations reproduce this B(E2)
value. The predicted values are orders of magnitude smaller  than
the  experimental  value.  These theoretical values correspond to
longer half life for this state, ranging from around 10 ns (B(E2)
=19.8  $e^2fm^4$) to 30 ns (B(E2) = 6.6 $e^2fm^4$). In the present
work,   definite   experimental  evidences  have  been  discussed
(Section IIIB) to indicate a longer half life  ($\simeq$  50  ns)
for  this  state.  The  theoretical  results  also  support  this
estimation from experimental observations.

\section{Summary   and   Conclusion}

The  high-spin  states  in  $^{153}$Ho,  have  been  populated by
$^{139}_{57}La$($^{20}Ne$, 6n) reaction at a projectile energy of
139 MeV at Variable  Energy  Cyclotron  Centre  (VECC),  Kolkata,
India.  An earlier campaign of Indian National Gamma Array (INGA)
setup has been utilized in this experiment. Data from gamma-gamma
coincidence,    directional    correlation    and    polarization
measurements  have  been analyzed to assign and confirm the spins
and parities of the levels. We have also  utilized  the  RF-gamma
time  spectrum  to  investigate  the  isomers  in  the excitation
spectrum. A few additions and revisions  of  the  reported  level
scheme of $^{153}$Ho have been suggested. Lifetime of a high spin
isomer has been suggested to be longer than the earlier result.

The alignment plots of the excitation spectra have been useful to
understand  the different modes of excitation phenomenologically.
Theoretical  calculations,  both  TRS  and  PRM   have   provided
additional microscopic insight.

The   regularity  in  the  increase  in  alignment  with  angular
frequency for low energies for both positive and negative  parity
states  indicate mild collectivity. This is also supported by TRS
results  showing  prolate  minimum  for  low  angular  frequency.
Although  PRM calculations indicate similar agreement with energy
values for both prolate and oblate options, the  isomer  lifetime
at $31/2^+$ is best reproduced by a prolate deformation.

For  positive  parity this regularity is repeated till relatively
higher spins with turns in between which may  indicate  intrinsic
configuration  changes.  The  PRM calculations reproduce $43/2^+$
lifetime by an oblate option  and  the  $61/2^+$  lifetime  by  a
prolate option.

The  shape  coexistence  is clearly manifested in positive parity
alignment plot with a smooth branch (non-yrast) and  a  irregular
branch  (yrast)  beyond  $45/2^+$.  This is supported by PRM. The
results for prolate deformation with $new core$ agree  well  with
the  non-yrast branch. This shape coexistence is also observed in
TRS plots.

 The  higher  spin negative parity states are erratic as shown in
alignment plot. It also agrees well with PRM results  for  oblate
deformation and supported by oblate minimum in TRS plot.

Future   investigations   are  needed  to  improve  the  lifetime
measurements to confirm the conclusions from this work on  firmer
ground.

\section*{Acknowledgment}

The  authors would like to acknowledge all INGA collaborators for
their sincere help and cooperation. The authors  sincerely  thank
P.  K. Das (SINP) and the target laboratory of VECC, Kolkata, for
preparation  of  the  target.  Special  thanks  are  due  to  the
Cyclotron  staff of VECC for providing necessary beams during the
experiment. This work is an outcome of the Collaborative Research
Scheme  (CRS)  project  of  UGC-DAE  Consortium  for   Scientific
Research,  Kolkata Centre. One of the authors (D.P.) acknowledges
financial support from UGC-DAE-CSR-KC and IIEST, Shibpur,  Howrah
for  providing  research  fellowship  at different stages of this
work.


\begin{thebibliography}{99}
\bibitem{nndc} Data extracted using the NNDC On-line Data Service
from the ENSDF database version of March 25, 2016.
\bibitem{shape} Kris Heyde, John L. Wood, Rev. of Mod. Phys, {\bf 83}, 1467 (2011).
\bibitem{dy152a} M.A. Bentley {\it et al.}, J. Phys. G: Nucl. Part. Phys. {\bf 17}, 481 (1991) and references therein.
\bibitem{dy152b} M. B. Smith {\it et al.}, Phys. Rev. C {\bf  61}, 034314 (2000) and references therein.
\bibitem{iso} J. Jastrzebski {\it et al.}, Phys. Lett. B {\bf 97}, 50 (1980).
\bibitem{iso1} Ani Aprahamian and Yang Sun, Nature Physics {\bf 1}, 81 (2005) and references therein.
\bibitem{ho151} C.T. Zhang {\it et al.}, Z. Phys. A {\bf 348}, 65 (1994).
\bibitem{ho152a}M. A. Rizzutto {\it et al.}, Phys. Rev. C {\bf 55}, 1130 (1997).
\bibitem{ho152b}S. Andre {\it et al.}, Nucl. Phys. A {\bf 575}, 155 (1994).
\bibitem{153hon} R. G. Helmer, Nucl. Data Sheets {\bf 107}, 507 (2006).
\bibitem{ho154} S.J. Chae {\it et al.}, Z. Phys. A {\bf 350}, 89 (1994).
\bibitem{dey} G. Dey {\it et al.}, Proc. DAE-BRNS Symp. Nucl. Phys. (India) {\bf 51}, 284 (2006), http://www.sympnp.org/proceedings/.
\bibitem{anagha} A.Chakraborty {\it et al.}, Proc. DAE-BRNS Symp. Nucl. Phys. {\bf 53}, 249(2008), http://www.sympnp.org/proceedings/.
\bibitem{db1} Dibyadyuti Pramanik {\it et al.}, Proc. DAE-BRNS Symp. Nucl. Phys. {\bf 55},14 (2010), http://www.sympnp.org/
proceedings/.
\bibitem{db2}Dibyadyuti Pramanik {\it et al.}, Proc. DAE-BRNS Symp. Nucl. Phys. {\bf 55},74 (2010), http://www.sympnp.org/
proceedings/.
\bibitem{db3}Dibyadyuti Pramanik {\it et al.}, Proc. DAE-BRNS Symp. Nucl. Phys.  {\bf 56}, 392 (2011), http://www.sympnp.org/
proceedings/.
\bibitem{db4}Dibyadyuti Pramanik {\it et al.}, Proc. DAE-BRNS Symp. Nucl. Phys.  {\bf 57}, 212 (2012), http://www.sympnp.org/
proceedings/.
\bibitem{db5}Dibyadyuti Pramanik {\it et al.}, Proc. DAE-BRNS Symp. Nucl. Phys. {\bf 58}, 302 (2013), http://www.sympnp.org/
proceedings/.
\bibitem{153ho} D.C. Radford {\it et al.}, Phys. Lett. B {\bf 126}, 24 (1983).
\bibitem{153hotri} D. E. Appelbe {\it et al.}, Phys. Rev. C {\bf 66}, 044305 (2002).
\bibitem{mom} G.D. Alkhazov {\it et al.}, Nucl. Phys.A {\bf 504}, 549 (1989).
\bibitem{raut} R. Raut {\it et al.}, Proc. DAE-BRNS Symp. Nucl. Phys. (India) {\bf 47B}, 578 (2004),http://www.sympnp.org/
proceedings/.
\bibitem{ingasort}  Ranjan K. Bhowmik in
{\it Structure of atomic nuclei}, proceedings of the SERC School, Puri, India,  1996,
edited by L. Satpathy (Narosa Publishing House, New Delhi, 1999) p. 259.
\bibitem{radware}  D. C. Radford, Nucl. Inst. Meths. A {\bf 361}, 297 (1995).
\bibitem{macias}  E.S.  Macias,  W.D. Ruhter, D.C. Camp, R.G. Lanier, Comput. Phys. Commun. {\bf 11}, 75 (1976).
\bibitem{30p} Indrani Ray {\it et al.}, Phys. Rev. C {\bf  76}, 034315 (2007).
\bibitem{34cl} Abhijit Bisoi {\it et al.}, Phys. Rev. C {\bf  89}, 024303 (2014).
\bibitem{33s} Abhijit Bisoi {\it et al.}, Phys. Rev. C {\bf  90}, 024328 (2014).
\bibitem{142sm} S  Rajbanshi {\it et al.}, Phys. Rev. C {\bf 89}, 014315 (2014).
\bibitem{bricc}T. Kibedi, T.W. Burrows, M.B. Trzhaskovskaya, P.M. Davidson, C.W. Nestor, Jr.,
Nucl. Instr. and Meth. A {\bf 589}, 202 (2008), http://bricc.anu.edu.au/index.php.
\bibitem{trs} W. Nazarewicz, R. Wyss and A. Johnson,  Nucl. Phys.  A {\bf 503}, 285 (1989).
\bibitem{prm1}E. M. M\"uller and U. Mosel, J. Phys. G {\bf 10}, 1523 (1984).
\bibitem{prm2}  M. Saha, A. Goswami,  S. Bhattacharya, S. Sen, Phys. Rev. C {\bf 42}, 1386 (1990).
\bibitem{nil}S. G. Nilsson, C. F. Tsang, A. Sobiczewski, Z. Szymanski,
S.Wycech, C. Gustafson, I. L. Lamm, P. Moller, and B. Nilsson,
Nucl. Phys. A {\bf 131}, 1 (1969).
\bibitem{raman}  S. Raman, C. W. Nestor, Jr., and P. Tikkanen, At. Data and Nucl. Data Tables {\bf 78}, 1 (2001).
\bibitem{audi} G. Audi {\it et al.,} Chinese Physics C {\bf  36}, 1157 (2012).



\end{thebibliography}
\end{document}